\documentclass[aps,prd,showpacs,eqsecnum,twocolumn]{revtex4}
\usepackage{amsmath,amssymb}
\usepackage{graphicx}
\usepackage{dcolumn}
\usepackage{bm}
\usepackage{epsf}

\draft

\newcommand{\beq}{\begin{equation}}
\newcommand{\eeq}{\end{equation}}
\newcommand{\beqn}{\begin{eqnarray}}
\newcommand{\eeqn}{\end{eqnarray}}

\def\agt{\mathrel{\raise.3ex\hbox{$>$}\mkern-14mu\lower0.6ex\hbox{$\sim$}}}
\def\alt{\mathrel{\raise.3ex\hbox{$<$}\mkern-14mu\lower0.6ex\hbox{$\sim$}}}

\begin{document}

\title{Stably stratified stars containing magnetic fields whose toroidal
  components are much larger than the poloidal ones in general relativity
  -- A perturbation analysis --}

\author{Shijun Yoshida$^{1}$
\footnote{\affiliation\ yoshida@astr.tohoku.ac.jp}}
\affiliation{$^{1}$Astronomical Institute, Tohoku University, Sendai 980-8578, Japan~}

\date{\today}

\begin{abstract}

We construct the stably stratified magnetized stars 
within the framework of general relativity. The effects 
of magnetic fields on the structure of the star and 
spacetime are treated as perturbations of  
non-magnetized stars. By assuming ideal 
magnetohydrodynamics and employing 
one-parameter equations of state, we derive 
basic equations for describing stationary and 
axisymmetric stably stratified stars containing  
magnetic fields whose toroidal components are much 
larger than the poloidal ones. A number of the polytropic 
models are numerically calculated to investigate 
basic properties of the effects of magnetic fields on 
the stellar structure. According to the stability result 
obtained by Braithwaite, which remains a matter of 
conjecture for general magnetized stars, certain of 
the magnetized stars constructed in this study 
are possibly stable.  

\end{abstract}

\pacs{04.40.Dg, 97.60.Jd}

\maketitle

\section{Introduction}\label{sec:Intro}

It has been well-accepted that soft-gamma repeaters (SGRs) and 
anomalous x-ray pulsars (AXPs) are magnetars, highly magnetized 
neutron stars whose strength of the surface field is as large as 
$\sim 10^{14}-10^{15}$~G~\cite{duncan,paczynsk,thompsona,thompsonb,woods}.
The existence of the magnetar has reactivated studies on equilibrium 
configurations of magnetized stars.

In order to elucidate basic properties of equilibrium configurations of 
magnetized stars, a large number of studies have been performed 
so far since the pioneering work of Chandrasekhar and Fermi~\cite{chandra}. 
A large fraction of those studies have been done within the framework of 
Newtonian magnetohydrodynamics and Newton's theory of gravity (cf., e.g., 
Refs.~\cite{prendergast,woltjer,roxburgh,trehanb,monaghana,monaghanb,
sinha,trehana,ioka,miketinac,miketinac2,tomimura,yoshida0,yoshida,
lander,fujisawaa,fujisawab,duez}). Since neutron stars are very compact 
in the sense that their compactness $M/R$ is as large as $\sim 0.1-0.2$ 
with $M$ and $R$ being their mass and radius in geometrical units, general 
relativity is required to describe the gravitational field of neutron stars. 
Therefore, general relativistic models of magnetized stars have been 
investigated as well. Bocquet et al.~\cite{bocquet} and Cardall et 
al.~\cite{cardall} obtained relativistic neutron star models with purely poloidal 
magnetic fields. Using a perturbative technique, Konno et al.~\cite{konno} 
calculated similar models to those obtained in Refs.~\cite{bocquet,cardall}. 
Kiuchi and Yoshida~\cite{kiuchi} computed 
magnetized stars with purely toroidal fields (cf., also, Ref.~\cite{frieben}). 
Ioka and Sasaki~\cite{Ioka2004}, Colaiuda et al.~\cite{colaiuda}, and Ciolfi 
et al.~\cite{ciolfia,ciolfib} derived relativistic stellar models having both 
toroidal and poloidal magnetic fields with perturbative techniques (cf., also, 
Ref.~\cite{ciolfic}). Yoshida et al.~\cite{yoshidaa} included the effects of 
the stable stratification in the magnetized star model obtained in Ref.~\cite{Ioka2004}. 
Uryu et al.~\cite{uryua,uryub} obtained magnetized stars with mixed poloidal-toroidal 
magnetic fields by solving a full set of Einstein equations, magnetohydrodynamics 
equations, and coordinate conditions numerically. By assuming 
simpler conformally flat spacetime, Pili et al.~\cite{pilia,pilib,bucciantini,pilic} 
calculated many models of magnetized stars. Although great  progress has 
been achieved in this field, as mentioned before, further studies are required 
because all the magnetized star models are constructed by some particular 
magnetic-field configurations which are not necessarily realistic. In 
particular, it is not still clear at all whether stable models exist. 

Stability of magnetized stars with a relatively simple magnetic field 
structure have been examined with analytical approaches. The pioneering 
work was done by Tayler~\cite{tayler73}, who showed that stars with purely 
toroidal magnetic fields are unstable. Wright~\cite{wright73} subsequently 
showed that the same instability mechanism, the pinch-type instability 
mechanism, operates in stars with purely poloidal magnetic fields. He also
suggested the possibility that stars having mixed poloidal-toroidal
magnetic fields may be stable if the strength of both components is
comparable (cf., also, Refs.~\cite{markey,tayler80,assche}). 
Flowers and Ruderman~\cite{flowers} found that another type of
instability occurs in purely poloidal magnetic field configurations.
All those classical stability analyses were based on a method of
an energy principle in the framework of Newtonian dynamics (cf., also,
Refs.~\cite{pitts,goossens}).  Another approach is a local analysis,
with which Acheson \cite{acheson1978} investigated the stability of
rotating magnetized stars containing purely toroidal fields in detail
within the framework of Newtonian dynamics (cf., also,
Refs.~\cite{acheson1979,spruit}) and derived detailed stability
conditions for purely toroidal magnetic fields buried inside rotating
stars with dissipation. Bonanno and Urpin analyzed the axisymmetric 
stability~\cite{bonanno} and the non-axisymmetric stability~\cite{bonanno2} 
of cylindrical equilibrium configurations possessing mixed poloidal-toroidal 
fields, while ignoring compressibility and stratification of the fluid.

Recently the stability problem of the magnetized star has been
approached from another direction, dynamical simulation approaches, 
and some significant progress has been made. By following the time 
evolution of small random initial magnetic fields around a spherical 
star in the framework of Newtonian resistive magnetohydrodynamics, 
Braithwaite and Spruit~\cite{Braithwaite2004,Braithwaite2006} 
obtained stable equilibria of magnetized stars that are formed as 
a self-organization phenomenon.  The resulting stable magnetic 
fields have both poloidal and toroidal components with comparable 
strength and support the conjecture for stability conditions of the 
magnetized star given by the classical studies mentioned before 
(cf., also, Ref.~\cite{duez10}). By using the numerical 
magnetohydrodynamic simulation, Braithwaite \cite{Braithwaite2009} 
studied stability conditions for the magnetized stars and obtained 
a stability condition for his models given in terms of the ratio of 
the poloidal magnetic energy to the total magnetic energy. The stability 
condition is given by  
\beq
\tilde{a} {E_{\rm EM}\over |W|} <  {E_{\rm EM}^{(p)} \over E_{\rm EM}} 
\lesssim 0.8 \,, 
\label{Braithwaite_SC}
\eeq
where $E_{\rm EM}$, $E_{\rm EM}^{(p)}$, and $W$ are the 
total magnetic energy, the poloidal magnetic energy, and the 
gravitational energy, respectively, and $\tilde{a}$ is a 
dimensionless factor related to the buoyancy properties 
of the star. For neutron stars and main-sequence stars, the 
dimensionless factor $\tilde{a}$ is of order $10^3$ and $10$, 
respectively. Lander and Jones examined the stability of 
magnetized stars by numerically solving the time evolution of 
linear perturbations of the stars in their series of 
papers~\cite{lander11,lander11b,lander12}.  For the stars with 
purely toroidal and purely poloidal magnetic fields, their results 
are consistent with those of the classical stability analysis, i.e., 
the pinch-type instability occurs near the symmetry and the 
magnetic axes for the cases of the purely toroidal and the purely 
poloidal magnetic fields, respectively (cf., also, 
Refs.~\cite{kiuchia,kiuchib,ciolfid,lasky,ciolfie,zink}). They also 
assessed the stability of various magnetized stars with mixed 
poloidal-toroidal fields and found that all their models considered 
suffer from the pinch-type instability even for the cases in which 
the poloidal and toroidal components have comparable 
strength~\cite{lander12}. At first glance, it seems that the results 
by Lander and Jones are incompatible with those by
Braithwaite and his collaborators~\cite{Braithwaite2009,duez10}. 
Mitchell et al.~\cite{mitchell} made numerical simulations smilar 
to those of Braithwaite and his 
collaborators~\cite{Braithwaite2004,Braithwaite2006} but for 
the case of the non-stratified star. They then obtained 
no stable equilibrium for the non-stratified case and showed 
that {\em stable stratification of the fluid} will be a key ingredient, 
which is taken into account in the analyses of 
Refs.~\cite{Braithwaite2009,duez10} but not in the analyses 
of Ref.~\cite{lander12}. In other words, the results obtained by 
Mitchell et al.~\cite{mitchell} suggest that stable stratification is 
required to avoid instability for some magnetic-field configurations 
inside the star. Note that in the simulations by Braithwaite 
and his collaborators~\cite{Braithwaite2004,Braithwaite2006}, 
resistive dissipation will also play a crucial role. Therefore, 
effects of the resistive dissipation on dynamical stability of 
the magnetized star need to be closely examined. 

Despite the fact that a large number of studies on equilibria 
and stabilities of magnetized stars have been made so far, 
as mentioned before, the magnetic field structure of the 
neutron star has not yet been elucidated not only theoretically 
but also observationally. The formation process of the neutron 
star would however provide us with some clues. During the core 
collapse events which produce neutron stars, the poloidal 
magnetic field lines would get wrapped around the rotation 
axis because of the differential rotation of the core (cf., e.g., 
Ref.~\cite{kotake}). As a result, the toroidal field would be 
significantly amplified. It is therefore likely to expect that the 
toroidal component of the magnetic field is much larger than 
the poloidal one inside the neutron star at least soon after its 
birth. 

To investigate properties of the magnetized star whose toroidal 
fields are much larger than the poloidal ones, Kiuchi and 
Yoshida~\cite{kiuchi} constructed the magnetized stars 
completely neglecting the poloidal component of the magnetic 
field. Although studies on stars with purely toroidal magnetic 
fields can elucidate approximate properties of magnetized stars 
whose toroidal fields are much larger than the poloidal ones, 
purely toroidal magnetic fields inside the star are unstable as 
mentioned before. To stabilize the 
toroidal magnetic field inside the star, the inclusion of the 
poloidal magnetic field is necessary. To our knowledge, however, 
equilibrium states of the magnetized star whose toroidal fields  
are much larger than the poloidal ones, which are plausible 
neutron star models, have not been constructed 
so far, except the case of purely toroidal magnetic fields. 
As mentioned before, another important stabilizing agent for 
magnetic fields inside the star is a stable stratification of the 
fluid. In order to construct neutron star models with more 
realistic interior magnetic field structure, 
in this study, we investigate stably stratified stars 
having magnetic fields characterized by the condition of 
${E_{\rm EM}^{(p)}/ E_{\rm EM}} \ll 1$, i.e., the toroidal field is  
much larger than the poloidal one, within the framework 
of general relativity. 

The strength of the effects of magnetic 
fields on the stellar structure can be roughly estimated by an 
approximate ratio of the magnetic field energy, $E_{\rm EM}$, 
to the gravitational energy, $W$, given by 
\beq
{E_{\rm EM}\over |W|} 
\approx 10^{-6} \left({B_0 \over 10^{15} \, {\rm G}}\right)^2 
\left({R\over 10 \, {\rm km}}\right)^4 \left({M\over 1.4 \,  M_\odot } \right)^{-2} \,,
\label{E/W_estimate}
\eeq
where $B_0$, $R$, and $M$ are the strength of the magnetic 
field, the radius, and the mass of the star, respectively. This 
ratio is very small even if a magnetar characterized by 
$B_0 \sim 10^{15} \, {\rm G}$ is considered. In order to 
investigate effects of magnetic fields on the neutron star 
structure, therefore, perturbation approaches are generally 
quite efficient in the sense that they are tractable and give 
sufficiently accurate results. We therefore make use of a 
perturbation approach to study the structure of the 
magnetized star in this work. 

The present paper is organized as follows. In 
Section~\ref{sec:basic_eq}, we introduce the general 
formalism for general relativistic ideal magnetohydrodynamics. 
Section~\ref{sec:formulation} presents the formalism used to 
construct the stably stratified magnetized star whose toroidal 
fields are much larger than the poloidal ones. In 
Section~\ref{sec:result}, we exhibit  examples of the stably 
stratified magnetized stars calculated numerically. Finally, we 
give the discussion and summary in 
Sections~\ref{sec:discussion} and \ref{sec:summary}, 
respectively. In Appendix, we give a Newtonian analysis 
of the same magnetized star as that discussed in this paper. 
In the following, we choose the signature $(-,+,+,+)$ for 
the spacetime metric and, unless otherwise stated, we 
adopt geometrical units with $c=G=1$, where $c$ and 
$G$ are the speed of light and Newton's gravitational 
constant, respectively. 

\section{Basic equations describing dynamics of perfectly 
conductive fluids}\label{sec:basic_eq}

The dynamics of perfect fluids coupled with electromagnetic fields 
may be described by the magnetohydrodynamics equations 
summarized as follows. Baryon mass conservation equation: 
\beq
\nabla_\mu\left(\rho u^\mu\right)=0\,, 
\label{baryon} 
\eeq
where $\rho$ and $u^\mu$ are the rest-mass density and 
the fluid four-velocity, respectively. Here, $\nabla_\mu$ 
denotes the covariant derivative associated with the metric 
$g_{\mu\nu}$, and spacetime indices are denoted by lower 
case Greek letters ($\alpha, \beta, \gamma, \cdots$). The 
Maxwell equations: 
\beqn
&&\nabla_\alpha F_{\mu\nu}+\nabla_\mu F_{\nu\alpha}
+\nabla_\nu F_{\alpha\mu}=0\,, 
\label{maxwell1}
\\
&& \nabla_\nu F^{\mu\nu}=4\pi J^\mu\,, 
\label{maxwell2}
\eeqn
where $F_{\mu\nu}$ and $J^\mu$ are the Faraday tensor 
and the current four-vector, respectively. The conservation 
law of the energy-momentum tensor: 
\beq
\nabla_\nu T^{\mu\nu}=0\,,
\label{devT}
\eeq
where $T^{\mu\nu}$ is the energy-momentum tensor, 
defined by 
\beqn
T^{\mu\nu}&=&\rho h u^\mu u^\nu + P g^{\mu\nu} \nonumber \\ 
&& + {1\over 4\pi}\left[ F^{\mu\alpha}{F^\nu}_\alpha 
- {1\over 4}g^{\mu\nu}F^{\alpha\beta}F_{\alpha\beta}
\right]\,,
\label{Total_T}
\eeqn
where $h$ and $P$ are the specific enthalpy and the 
pressure, respectively. Here, the specific enthalpy  may, 
in terms of the specific internal energy $\varepsilon$, the 
pressure $P$, and the rest-mass density $\rho$, be 
defined by 
\beq
h = 1+\varepsilon+{P\over \rho}\,. 
\label{Def_h}
\eeq
As for the equations of state, we supply one-parameter 
equations of state, given by 
\beq
P=P\left(\rho\right)\,, \quad \varepsilon=\varepsilon\left(\rho\right) \,. 
\label{Def_one_p_EOS}
\eeq
The electric field $E_\mu$ and the magnetic field $B_\mu$ 
observed by an observer associated with the fluid 
four-velocity $u^\mu$ are defined by 
\beqn
E_\mu&=&F_{\mu\nu}u^\nu\,,\\
B_\mu&=&{1\over 2}\epsilon_{\nu\mu\alpha\beta}u^\nu F^{\alpha\beta}\,,
\eeqn
where $\epsilon_{\mu\nu\alpha\beta}$ is the Levi-Civita 
tensor with $\epsilon_{0123}=\sqrt{-g}$. Here, $g$ denotes 
the determinant of the metric $g_{\mu\nu}$. Since the neutron-star 
matter may be approximately assumed as a perfect conductor, 
in this study, we may further impose the condition of perfect 
conductivity, given by 
\beq
E_\mu=F_{\mu\nu}u^\nu=0\,. 
\label{MHD_condition}
\eeq
Equation (\ref{devT}) may be divided into two equations, the 
energy equation and the momentum equations, respectively, 
given by 
\beqn
-u_\mu \nabla_\nu T^{\mu\nu}&=&u^\mu\nabla_\nu\left\{\rho\left(1+\varepsilon\right)\right\}+\rho h \nabla_\nu u^\nu
\nonumber \\ 
&=&\rho u^\nu\nabla_\nu\varepsilon+P\nabla_\nu u^\nu=0\,,
\label{energy_eq}\\
q_{\mu\alpha}\nabla_\nu T^{\alpha\nu}&=&\rho h u^\nu\nabla_\nu u_\mu+q_\mu^\nu\nabla_\nu P
-F_{\mu\nu}J^\nu \nonumber \\
&=&0\,, 
\label{momentum_eq}
\eeqn
where $q_{\mu\nu} = g_{\mu\nu}+u_\mu u_\nu$. Note 
that the perfect conductivity condition (\ref{MHD_condition}) 
has been used in the derivation of Equation (\ref{energy_eq}). 

\section{Master equations for equilibrium solutions of the 
magnetized star}\label{sec:formulation}

In order to obtain equilibrium solutions of the relativistic stars 
containing the mixed poloidal-toroidal magnetic fields, in this 
study, we make the following assumptions: (i) Equilibrium 
models are stationary and axisymmetric, i.e., the spacetime 
has the time Killing vector $t^\mu$ and the rotational Killing 
vector $\varphi^\mu$, and  Lie derivatives of the equilibrium 
quantities along the Killing vectors $t^\mu$ and $\varphi^\mu$ 
vanish. (ii) There is no fluid flow. (iii) The magnetic fields are 
sufficiently weak in the sense that the magnetic effects on the 
equilibrium structures may be treated as perturbations of  
stars including no electromagnetic field. (iv) The toroidal 
component of magnetic fields is much larger than the poloidal 
one. Under these assumptions, we may derive the master 
equations for describing equilibrium states of the relativistic 
stars containing the mixed poloidal-toroidal magnetic fields 
using the magnetohydrodynamic equations summarized in 
the previous section. 

In order to give a clear and  definite description of the 
assumptions~(iii) and (iv), we introduce two dimensionless 
smallness parameters $\varepsilon_t$ and $\varepsilon_p$ 
representing the amplitudes of the toroidal and the poloidal 
components of magnetic fields, respectively. We may then 
write that 
$^{(t)}F_{\mu\nu} { }^{(t)}F^{\mu\nu}\, P^{-1} = O\left(\varepsilon_t^2\right)$ 
and  
$^{(p)}F_{\mu\nu}{ }^{(p)}F^{\mu\nu}\, P^{-1} = O\left(\varepsilon_p^2\right)$ 
where $^{(t)}F_{\mu\nu}$ and $^{(p)}F_{\mu\nu}$ stand for the 
toroidal and poloidal components of the Faraday tensor 
$F_{\mu\nu}$, respectively. Note that $F_{\mu\nu}$ can be 
divided into the two parts, $^{(t)}F_{\mu\nu}$ and 
$^{(p)}F_{\mu\nu}$, because of the assumptions of the stationary 
and axially symmetric magnetic fields and the perfectly conducting 
fluid without flow. By the assumption~(iii) we have 
$\varepsilon_t\ll 1$ and $\varepsilon_p\ll 1$. By the 
assumption~(iv) we further impose that 
$\varepsilon_p\ll \varepsilon_t\ll 1$. 

In this study, as mentioned before, the unperturbed state is 
assumed to be a static and spherically symmetric star without 
magnetic fields. Around the spherically symmetric star, we may 
impose as perturbations the magnetic fields, given by 
\begin{eqnarray}
F_{\mu\nu}=\varepsilon_t { }^{(t)}F_{\mu\nu} + \varepsilon_p { }^{(p)}F_{\mu\nu} \,.
\end{eqnarray}
By this magnetic field, the matter distribution deviates from 
spherical symmetry. In this study, we are primarily interested 
in the lowest-order effects of the poloidal magnetic field on 
the structure of the star including purely toroidal magnetic fields. 
We therefore consider perturbations of order $\varepsilon_t^2$ 
and $\varepsilon_t \varepsilon_p$ on the structure of the 
spherically symmetric star but neglect perturbations of order 
higher than $\varepsilon_p^2$. Note that because of the 
assumption~(iv), i.e., $\varepsilon_p\ll \varepsilon_t\ll 1$, we 
have the inequality  
$
\varepsilon_t^0 \varepsilon_p^0 \gg \varepsilon_t^2 \varepsilon_p^0 
\gg \varepsilon_t^1 \varepsilon_p^1 \gg \varepsilon_t^0 \varepsilon_p^2\,. 
$

\subsection{Static and spherically symmetric stars without 
magnetic fields:  the $\varepsilon_t^0 \varepsilon_p^0$-order 
equations}\label{subsec:unperturbed_eq}
The line element of static and spherically symmetric spacetime 
may be given by 
\begin{eqnarray}
ds^2 &=& ^{(0)}g_{\mu\nu}dx^\mu dx^\nu \\
&=&-e^{2\nu}dt^2+e^{2\lambda}dr^2 \nonumber \\
&&\quad \quad +r^2\left( d\theta^2+\sin^2\theta\,d\varphi^2 \right) \,,
\label{eq:metric0}
\end{eqnarray}
where $^{(0)}g_{\mu\nu}$ denotes the unperturbed metric, 
and $\nu$ and $\lambda$ are functions of $r$ only. The equilibrium state 
of the unperturbed star is described by the following equations 
(cf., e.g., Ref.~\cite{mtw}):
\beqn
&&{dM_r\over dr}=4\pi r^2 {\,}^{(0)}\rho\left(1+{\,}^{(0)}\varepsilon\right)\,,\label{eq:enclose}\\
&&{d{\,}^{(0)}P\over dr}=-e^{2\lambda}{\,}^{(0)}\rho {\,}^{(0)}h{M_r+4\pi {\,}^{(0)}P r^3 \over r^2}\,,\\
&&{d\nu\over dr}=-{1 \over {\,}^{(0)}\rho {\,}^{(0)}h}{d{\,}^{(0)}P\over dr}\,, 
\label{eq_dnudr}
\eeqn
where $M_r$ is defined in terms of the metric function by 
\beq
M_r = {r \over 2}\left(1-e^{-2\lambda}\right)\,, 
\eeq
and ${\,}^{(0)}\rho$, ${\,}^{(0)}\varepsilon$, ${\,}^{(0)}P$, and 
${\,}^{(0)}h$ are, respectively, the rest-mass density, specific 
internal energy, pressure, and specific enthalpy for the 
unperturbed star.

\subsection{Magnetic fields around a spherical star:  the  
$\varepsilon_t^1 \varepsilon_p^0$ and $\varepsilon_t^0 \varepsilon_p^1$ 
order equations}\label{subsec:mag_eq}
Due to the assumption of no fluid flow, the assumption~(ii), the fluid 
four-velocity is given by 
\begin{eqnarray}
u^\mu  = \gamma\,t^\mu\,, \label{eq:4-vel}
\end{eqnarray}
where $\gamma$ is the  function determined by the normalization 
condition $u^\mu u_\mu=-1$. The perfect-conductivity 
condition~(\ref{MHD_condition}) then becomes  
\begin{eqnarray}
F_{\mu\nu} t^\nu=F_{\mu t}= 0\,.\label{eq:MHD0}
\end{eqnarray}
As argued by Kiuchi and Yoshida~\cite{kiuchi}, the toroidal component 
of the magnetic field may be characterized by the conditions, given by 
\begin{eqnarray}
^{(t)}F_{\mu\nu} \varphi^\nu=0 \,.\label{eq:tor_con}
\end{eqnarray}
Thus, we see that the non-zero component of the toroidal magnetic 
field $^{(t)}F_{\mu\nu}$ is $^{(t)}F_{r\theta}\left(=-^{(t)}F_{\theta r}\right)$ 
only. The poloidal component of the magnetic field may be given in term 
of the poloidal flux function $\Psi$, which is actually the $\varphi$ 
component of the vector potential $A_\mu$, i.e., $\Psi=A_\varphi$ 
(cf., e.g., Ref.~\cite{bocquet}). 
Under the present assumptions, therefore, the Faraday tensor may be 
given by 
\begin{eqnarray}
F_{\mu\nu}&=&\varepsilon_t  {}^{(t)}F_{\mu\nu}+\varepsilon_p  {}^{(p)}F_{\mu\nu} \nonumber \\ 
&=&
\varepsilon_t \, ^{(t)}F_{r\theta} \left(
\begin{array}{cccc}
 0 & 0 & 0 & 0 \\
 0 & 0 & 1 & 0 \\
 0 &-1& 0 & 0 \\
 0 & 0 & 0 & 0
\end{array}
\right) \nonumber \\
&&+
\varepsilon_p\left(
\begin{array}{cccc}
 0 & 0 & 0 & 0 \\
 0 & 0 & 0 &\partial_r \Psi \\
 0 & 0 &0 & \partial_\theta \Psi \\
 0 & -\partial_r \Psi  & -\partial_\theta \Psi & 0
\end{array}
\right)\,,\label{comp:faraday}
\end{eqnarray}
where $^{(t)}F_{r\theta}$ and $\Psi$ are functions of $r$ and $\theta$ 
only. Thanks to the introduction of the poloidal flux function $\Psi$, we 
see that the Faraday tenser (\ref{comp:faraday}) automatically satisfies 
one of the Maxwell equations (\ref{maxwell1}). The other Maxwell 
equation (\ref{maxwell2}) is used to determine the current 
four-vector $J^\mu$ in ideal magnetohydrodynamic theory. The 
explicit form of the current four-vector $J^\mu$ is, from equations 
(\ref{maxwell2}), (\ref{eq:metric0}) and (\ref{comp:faraday}), given by 
\begin{eqnarray}
&& J^t=0\,, \quad J^\varphi=O\left(\varepsilon_p\right)\,, \\
&& J^r=\varepsilon_t {1\over 4\pi e^{\nu+\lambda}r^2\sin\theta}
\partial_\theta\left(e^{\nu-\lambda}\sin\theta ^{(t)}F_{r\theta}\right) \nonumber \\
&& \quad\quad\quad\quad + O\left(\varepsilon_t^3\right) \,, \\
&& J^\theta=\varepsilon_t {-1\over 4\pi e^{\nu+\lambda}r^2\sin\theta}
\partial_r\left(e^{\nu-\lambda}\sin\theta ^{(t)}F_{r\theta}\right) \nonumber \\
&& \quad\quad\quad\quad + O\left(\varepsilon_t^3\right) \,.
\end{eqnarray}
This current four-vector $J^\mu$ is used to write the momentum 
equation (\ref{momentum_eq}) explicitly, which becomes, in the 
present situation, 
\begin{eqnarray}
-\partial_\mu \ln \gamma +\frac{1}{\rho h} \partial_\mu P 
- \frac{1}{\rho h}F_{\mu\nu} J^\nu = 0\,.\label{eq:Euler}
\end{eqnarray}
Because of equation (\ref{eq:MHD0}), we see that the time 
component of equation (\ref{eq:Euler}) is automatically satisfied. 
The toroidal component ($\varphi$ component) and the poloidal 
components ($r$ and $\theta$ components) of equation 
(\ref{eq:Euler}), respectively, lead  
\begin{eqnarray}
&&\varepsilon_t\varepsilon_p\partial_\theta\left(e^{\nu-\lambda}\sin\theta ^{(t)}F_{r\theta}\right)\,\partial_r \Psi 
\nonumber \\ && - 
\varepsilon_t\varepsilon_p\partial_r\left(e^{\nu-\lambda}\sin\theta ^{(t)}F_{r\theta}\right)\,\partial_\theta \Psi
\nonumber \\ &&
+O\left(\varepsilon_t^3\varepsilon_p\right)=0  \,,\label{eq:Euler-phi}\\
&&-\partial_C \ln \gamma +\frac{1}{\rho h}\partial_C P \nonumber \\ 
&&+\varepsilon_t^2{ ^{(t)}F_{r\theta}\over 4\pi ^{(0)}\rho ^{(0)}h e^{\nu+\lambda}r^2\sin\theta}
\partial_C\left(e^{\nu-\lambda}\sin\theta ^{(t)}F_{r\theta}\right) \nonumber \\ 
&&+O\left(\varepsilon_p^2\right) = 0\,,
\label{eq:Euler-poloidal}
\end{eqnarray}
where the index $C$ is used to denote poloidal indices and 
runs from 1($r$)  to 2($\theta$). Note that 
equations~(\ref{eq:Euler-phi}) and (\ref{eq:Euler-poloidal}) are 
the $\varepsilon_t\varepsilon_p$-order accurate expression of 
the momentum equation (\ref{eq:Euler}). The integrability 
conditions for equations~(\ref{eq:Euler-phi}) and 
(\ref{eq:Euler-poloidal}) require that 
\begin{eqnarray}
\Psi&=&\Psi\left[e^{\nu-\lambda}\sin\theta ^{(t)}F_{r\theta}\right]\,, \label{def:func_Psi} \\ 
e^{\nu-\lambda}\sin\theta ^{(t)}F_{r\theta}&=&K\left[^{(0)}\rho ^{(0)}h e^{2\nu}r^2\sin^2\theta\right]\,, 
\label{def:func_K}
\end{eqnarray}
where $K$ is an arbitrary function of 
$^{(0)}\rho ^{(0)}h e^{2\nu}r^2\sin^2\theta$. Equations  
(\ref{def:func_Psi}) and (\ref{def:func_K}) are only  the conditions 
that the magnetic fields have to satisfy. Therefore, the magnetic 
field distribution can be specified by the two arbitrary functions 
$\Psi=\Psi\left[w\right]$ and $K=K\left[w\right]$ with $w$ being 
$w = ^{(0)}\rho ^{(0)}h e^{2\nu}r^2\sin^2\theta$ 
as far as the corresponding magnetic field satisfies the physically 
reasonable boundary conditions.

\subsection{Deformation of the star and spacetime due to the 
magnetic field: the  $\varepsilon_t^2 \varepsilon_p^0$ and 
$\varepsilon_t^1 \varepsilon_p^1$ order equations}
\label{subsec:mag_eq2}
In this subsection, we derive the master equations for the deformation 
of the star and spacetime due to the magnetic field discussed in the 
previous subsection. Since the fluid four-velocity is proportional to the 
time Killing vector $t^\mu$, as given in equation (\ref{eq:4-vel}), the 
baryon mass conservation equation (\ref{baryon}) and the energy 
equation (\ref{energy_eq}) are automatically satisfied. We do not 
therefore need to consider them further. The only fluid equation that 
we have to consider is equation (\ref{eq:Euler-poloidal}). Because of 
equations~(\ref{Def_one_p_EOS}) and (\ref{def:func_K}), the 
momentum equation~(\ref{eq:Euler-poloidal}) has the first integral, 
given by 
\begin{eqnarray}
&&- \ln \gamma + \int \frac{dP}{\rho h} 
+ \varepsilon_t^2 \frac{1}{4\pi} \int \frac{K(w)}{w}\frac{dK}{dw}dw 
\nonumber \\
&&\quad = C+O\left(\varepsilon_p^2\right),\label{eq:Ber}
\end{eqnarray}
where $C$ is a constant of integration. This equation is sometimes 
called the equation of hydrostatic equilibrium.  From equation~(\ref{eq:Ber}), 
it is seen that the poloidal magnetic field does not affect the fluid 
distribution within the $\varepsilon_t\varepsilon_p$ order accuracy. 
As shown later, the poloidal magnetic field does affect the 
spacetime geometry, which is determined by the Einstein equations.  

The functions $\gamma$ and $\displaystyle \int \frac{dP}{\rho h}$ 
and the constant of integration $C$ may be expanded as follows:  
\begin{eqnarray}
\gamma&=&e^{-\nu}+ \varepsilon_t^2 {}^{(2)}\gamma(r,\theta)
+O\left(\varepsilon_p^2\right) \,, \label{Exp_gamma}\\
\int \frac{dP}{\rho h}&=&\int \frac{d ^{(0)}P}{ ^{(0)}\rho { } ^{(0)}h} 
+ \varepsilon_t^2 \frac{ ^{(2)}P(r,\theta)}{ ^{(0)}\rho { } ^{(0)}h} 
+O\left(\varepsilon_p^2\right) \,, \label{Exp_dP} \\
C&=& ^{(0)}C+ \varepsilon_t^2 {}^{(2)}C+O\left(\varepsilon_p^2\right) \,, 
\label{Exp_C}
\end{eqnarray}
where $^{(2)}\gamma$, $^{(2)}P$, and $^{(2)}C$ are perturbations 
of order $\varepsilon_t^2$. Substituting 
equations~(\ref{Exp_gamma})--(\ref{Exp_C}) into equation~(\ref{eq:Ber}), 
we obtain 
\begin{eqnarray}
&&\int \frac{d ^{(0)}P}{ ^{(0)}\rho { } ^{(0)}h} +\nu= ^{(0)}C\,,\label{eq:Ber0} \\
&& \frac{ ^{(2)}P(r,\theta)}{ ^{(0)}\rho { } ^{(0)}h} - e^{\nu} {}^{(2)}\gamma(r,\theta)
\nonumber \\ && \quad 
+ \frac{1}{4\pi} \int \frac{K(w)}{w}\frac{dK}{dw}dw = ^{(2)}C\,.\label{eq:Ber2}
\end{eqnarray}
Note that the derivative of equation~(\ref{eq:Ber0}) with respect to 
$r$ yields equation~(\ref{eq_dnudr}). 

The stress-energy tensor given in equation (\ref{Total_T}) is divided 
into the fluid part $^{(F)}T^\mu_\nu$ and the electromagnetic part 
$^{(EM)}T^\mu_\nu$, defined, respectively, by
\begin{eqnarray}
^{(F)}{T^\mu}_\nu&=&\rho h u^\mu u_\nu + P {\delta^\mu}_\nu \,, \\
^{(EM)}{T^\mu}_\nu&=&{1\over 4\pi}\left( 
F^{\mu\alpha}F_{\nu\alpha}-{1\over 4}{\delta^\mu}_\nu\, 
F^{\alpha\beta}F_{\alpha\beta}
\right) \,.
\end{eqnarray}
For the Faraday tensor given in equation (\ref{comp:faraday}), the 
electromagnetic part of the stress-energy tensor is given by 
\begin{eqnarray}
&&^{(EM)}{T^\mu}_\nu  \nonumber \\
&&=
\varepsilon_t^2\,{e^{-2\lambda}\over 8\pi r^2}\left(^{(t)}F_{r\theta}\right)^2\left(
\begin{array}{cccc}
 -1 & 0 & 0 & 0 \\
 0 & 1 & 0 &0 \\
 0 & 0 & 1 & 0 \\
 0 & 0 & 0 & -1
\end{array}
\right) \nonumber \\
&&+
\varepsilon_t\varepsilon_p{e^{-2\lambda}\,{}^{(t)}F_{r\theta}\over 4\pi r^2}\left(
\begin{array}{cccc}
 0 & 0 & 0 & 0 \\
 0 & 0 & 0 & - \partial_\theta \Psi  \\
 0 & 0 & 0 & \ \  \partial_r \Psi \\
 0 & \displaystyle-r^{-2} e^{2\lambda} {\partial_\theta \Psi \over \sin^2\theta} 
   &\displaystyle {\partial_r \Psi\over \sin^2\theta}  & 0
\end{array}
\right) \nonumber \\ 
&&+O\left(\varepsilon_p^2\right)\,. \label{T_EM_general}
\end{eqnarray}
From the expression (\ref{T_EM_general}), we may confirm that 
whereas the circularity conditions for the spacetime, given by 
\begin{eqnarray}
t^\alpha {T_\alpha}^{[\beta}t^\gamma \varphi^{\delta ]}=0\,, \quad 
\varphi^\alpha {T_\alpha}^{[\beta}t^\gamma \varphi^{\delta ]}=0\,, 
\label{circularity_C}
\end{eqnarray}
(cf., e.g., Ref.~\cite{wald}) are fulfilled up to the $\varepsilon_t^2$ 
order, they are violated at the $\varepsilon_t \varepsilon_p$ order. 
This implies that up to the $\varepsilon_t^2$ order, the spacetime 
around the magnetized stars considered may be described by a 
simpler form of the metric used for stationary and axisymmetric 
rotating stars without magnetic fields (cf., e.g., Ref.~\cite{thorne}). 
Note that the solutions within accuracy up to $\varepsilon_t^2$ 
order correspond to a perturbation version of the star containing  
purely toroidal magnetic fields constructed by Kiuchi and 
Yoshida~\cite{kiuchi}. 

To calculate particular models of the magnetized star, we need to 
specify completely the arbitrary functions $K$ and $\Psi$, given in 
equations~(\ref{def:func_Psi}) and (\ref{def:func_K}). In the 
present study, the two arbitrary functions are assumed to be 
given by 
\begin{eqnarray}
&&K=b\, w=b\,  ^{(0)}\rho ^{(0)}h e^{2\nu}r^2\sin^2\theta\,, \label{K} 
\label{def:F_K} \\
&& \Psi=a\, w=a\,  ^{(0)}\rho ^{(0)}h e^{2\nu}r^2\sin^2\theta \,, \label{Psi} 
\label{def:F_Psi}
\end{eqnarray}
where $b$ and $a$ are constants. Note that this choice of the 
function $K$ is the same as that of the $k=1$ case considered 
in Ref.~\cite{kiuchi}. For these arbitrary functions, the regularity 
of the magnetic field on the symmetry axis is satisfied. In this 
study, we assume that there is no magnetic field outside the 
star and that there is no surface current. Thus, the magnetic 
field has to vanish on the surface of the star. For the arbitrary 
functions given in equations~(\ref{def:F_K}) and 
(\ref{def:F_Psi}), the magnetic 
field $B^\mu$ becomes  
\begin{eqnarray}
B^\mu &=& \varepsilon_t  b \left(0, 0,0 ,  e^\nu  {}^{(0)}\rho  ^{(0)}h  \right) \nonumber \\
 &+&\varepsilon_p a\, e^{2\nu-\lambda} \left(0, 2\, {}^{(0)}\rho  ^{(0)} h \cos\theta , \nonumber \right. \\  
 &-& \left.  {\sin\theta\over r}\left\{ r {d\over dr}\left( ^{(0)}\rho  ^{(0)}h \right)
 +2 ( ^{(0)}\rho  ^{(0)}h) \left(r {d\nu\over dr}+1\right)\right\} , 0 \right)\nonumber \\
 &+&O\left(\varepsilon_t^3\right) \,. 
 \label{exp:B-field}
\end{eqnarray}
This magnetic field vanishes if the two conditions 
$^{(0)}\rho  ^{(0)}h =0$ and 
$\displaystyle {d\over dr}\left( ^{(0)}\rho  ^{(0)}h\right)=0$ 
are fulfilled. On the surface of the star, therefore, we require 
the conditions, given by 
\begin{eqnarray}
^{(0)}\rho  ^{(0)}h =0 \,, \quad {d\over dr}\left( ^{(0)}\rho  ^{(0)}h \right) =0 \,, 
\label{Condition:fluid}
\end{eqnarray}
which are, as a matter of fact, conditions for the equation of state.  

The explicit expression for non-zero components of 
$^{(EM)}{T^\mu}_\nu$ is summarized as follows:
\begin{eqnarray}
^{(EM)}{T^t}_t&=&^{(EM)}{T^\varphi}_\varphi=-^{(EM)}{T^r}_r=-^{(EM)}{T^\theta}_\theta \nonumber \\
&&=
-\varepsilon_t^2 \frac{b^2 }{8 \pi } r^2 e^{2 \nu} \left( ^{(0)}\rho  ^{(0)}h\right)^2 \sin^2\theta 
\nonumber \\ &&+O\left(\varepsilon_p^2\right)\,, 
\end{eqnarray}
\begin{eqnarray}
^{(EM)}{T^r}_\varphi&=&r^2\sin^2\theta \, e^{-2\lambda} {\,}^{(EM)}{T^\varphi}_r \nonumber \\
&=&
-\varepsilon_t\varepsilon_p\frac{a b }{2 \pi } r^2 e^{3 \nu-\lambda} \left( ^{(0)}\rho  ^{(0)}h\right)^2   \sin^2\theta  \cos\theta
\nonumber \\ &&+O\left(\varepsilon_t^3\right) \,, 
\end{eqnarray}
\begin{eqnarray}
^{(EM)}{T^\theta}_\varphi&=&\sin^2\theta {\,}^{(EM)}{T^\varphi}_\theta \nonumber \\
&=&\varepsilon_t\varepsilon_p\frac{a b}{4 \pi } r  e^{3 \nu-\lambda}  \left( ^{(0)}\rho  ^{(0)}h\right) 
\nonumber \\ &\times&
\left\{ r {d\over dr}\left( ^{(0)}\rho  ^{(0)}h \right)+2 ( ^{(0)}\rho  ^{(0)}h) \left(r {d\nu\over dr}+1\right)\right\}
\nonumber \\&\times&  \sin^3\theta +O\left(\varepsilon_t^3\right)\,. 
\end{eqnarray}
From this stress-energy tensor for the electromagnetic field, 
we may expect that the line element of the spacetime around 
the magnetized star is given by 
\begin{eqnarray}
ds^2 &=& - e^{2\nu}\left[1+2\epsilon_t^2 \left\{h_0(r)+h_2(r)P_2(\cos\theta)\right\}\right]dt^2 \nonumber \\
&&+ e^{2\lambda}\left[1+{2 \epsilon_t^2  e^{2\lambda}\over r}\left\{m_0(r)+m_2(r)P_2(\cos\theta)\right\}\right]dr^2 
\nonumber \\
&&+ r^2\left[1+2\epsilon_t^2 \left\{v_2(r)-h_2(r)\right\}P_2(\cos\theta)\right] \nonumber \\
&&\quad\quad \times \left(d\theta^2+ \sin^2\theta d\varphi^2\right)
\nonumber \\
&&-2\epsilon_t \epsilon_pd_2(r)\sin\theta{\partial\over\partial\theta}P_2(\cos\theta)\,dr d\varphi 
\nonumber \\ &&+O\left(\varepsilon_p^2\right)
\,,\label{eq:metric}
\end{eqnarray}
where $P_l$ is the Legendre polynomial of degree $l$. The 
normalization factor of the fluid four-velocity $\gamma$ 
defined by equation~(\ref{eq:4-vel}) (cf., also, 
equation~(\ref{Exp_gamma})) is then given by 
\begin{eqnarray}
\gamma&=&e^{-\nu}+ \varepsilon_t^2 \, {}^{(2)}\gamma(r,\theta)+O\left(\varepsilon_p^2\right)  \\
&=&e^{-\nu} \left[ 1 - \epsilon_t^2\left\{h_0(r)+h_2(r)P_2(\cos\theta)\right\} \right] \nonumber \\
&&+O\left(\epsilon_p^2\right) \,.\label{eq:gamma}
\end{eqnarray}
The $\epsilon_t^2$-order pressure perturbation given in 
equation~(\ref{eq:Ber2}) is written by  
\begin{eqnarray}
\frac{ ^{(2)}P(r,\theta)}{ ^{(0)}\rho { } ^{(0)}h} 
&=&\frac{ \delta P_0(r)}{ ^{(0)}\rho { } ^{(0)}h}+\frac{\delta P_2 (r) }{ ^{(0)}\rho { } ^{(0)}h}  P_2(\cos\theta) \nonumber \\
&=&^{(2)}C + e^{\nu(r)} {}^{(2)}\gamma(r,\theta) - \frac{1}{4\pi} \int \frac{K(w)}{w}\frac{dK}{dw}dw  \nonumber \\
&=&^{(2)}C - h_0(r)-\frac{1}{6\pi} b^2 {\,}^{(0)}\rho ^{(0)}  h \, e^{2\nu}r^2 \nonumber \\ 
&-& \left\{h_2(r)- \frac{1}{6\pi} b^2 {\,}^{(0)}\rho ^{(0)}  h \, e^{2\nu}r^2 \right\} P_2(\cos\theta) \,, \nonumber \\
\end{eqnarray}
where $\delta P_0$ and $\delta P_2$ are coefficients in the 
Legendre expansion of $^{(2)}P(r,\theta)$. Thus, we have
\begin{eqnarray}
\frac{ \delta P_0(r)}{ ^{(0)}\rho { } ^{(0)}h}  &=& -h_0(r)-\frac{1}{6\pi} b^2 {\,}^{(0)}\rho ^{(0)}  h \, e^{2\nu}r^2 
\nonumber \\ && + ^{(2)}C \,, \\
\frac{\delta P_2 (r) }{ ^{(0)}\rho { } ^{(0)}h} &=& -h_2(r)+ \frac{1}{6\pi} b^2 {\,}^{(0)}\rho ^{(0)}  h \, e^{2\nu}r^2  \,. 
\end{eqnarray}
From the relations obtained so far and the perturbed 
Einstein equations, following standard procedures (cf., 
e.g., Refs.~\cite{Ioka2004,yoshidaa}), we obtain the 
master equations for the deformation of the magnetized star with 
mixed poloidal-toroidal fields as follows.  

\noindent
The $\varepsilon_t^2$ order equations:
\begin{eqnarray}
{dm_0\over dr} &=&r^2 \left( ^{(0)}\rho { } ^{(0)}h\right) \nonumber \\ 
&&\times \left\{4\pi  \left({d ( ^{(0)}\rho + {\,}^{(0)}\rho {\,}^{(0)} \epsilon) \over d{\,}^{(0)} P}\right)\frac{ \delta P_0(r)}{ ^{(0)}\rho { } ^{(0)}h} 
\nonumber  \right. \\ && \left.  \quad\quad 
+\frac{1}{3}\,b^2 e^{2 \nu}r^2\left( ^{(0)}\rho { } ^{(0)}h\right)\right\}\,, \label{dm0_eq}\\
{d\over dr} \left( \frac{ \delta P_0(r)}{ ^{(0)}\rho { } ^{(0)}h} \right)&=&-\frac{e^{4 \lambda}}{r^2}  \left(1+8 \pi\,r^2{\,}^{(0)}P \right) m_0
\nonumber \\ &&
-4 \pi  e^{2 \lambda} r \left( ^{(0)}\rho { } ^{(0)}h\right) \left( \frac{ \delta P_0(r)}{ ^{(0)}\rho { } ^{(0)}h} \right) \nonumber \\
&&-b^2 e^{2\nu}\left\{\frac{r^2}{6 \pi }{d\over dr}\left(^{(0)}\rho { } ^{(0)}h\right)
 \right. \nonumber \\ && \left. 
  +\frac{r^3}{3} e^{2 \lambda} \left(^{(0)}\rho { } ^{(0)}h\right)\left(^{(0)}\rho { } ^{(0)}h+4{\,}^{(0)} P\right) \right. \nonumber \\
&&\left. +\frac{r}{6 \pi}\left(1+e^{2 \lambda}\right)\left(^{(0)}\rho { } ^{(0)}h\right)\right\}\,, \label{dp0_eq} \\
h_0&=&-\frac{ \delta P_0(r)}{ ^{(0)}\rho { } ^{(0)}h} -\frac{1}{6 \pi }\,{b^2 e^{2 \nu} r^2}\left(^{(0)}\rho { } ^{(0)}h\right) 
\nonumber \\ && + ^{(2)}C\,,
\end{eqnarray}
\begin{eqnarray}
{dh_2\over dr}&=&-\frac{2 e^{2 \lambda}}{\displaystyle r^2 {d\nu\over dr}}\,v_2 
\nonumber \\ &&
-\left[2 {d\nu\over dr}-\frac{e^{2 \lambda}}{\displaystyle r^3 {d\nu\over dr}}\left\{4 \pi r^3\left( ^{(0)} \rho ^{(0)} h\right)-2M\right\} \right]h_2
\nonumber \\
&&-\frac{b^2\,e^{2 \nu} }{\displaystyle 3 {d\nu\over dr}}\,r^2 \left(2 \left(r{d\nu\over dr}\right)^2+e^{2 \lambda}\right) \left( ^{(0)} \rho ^{(0)} h\right)^2 \,, 
\nonumber \\ 
 \label{dh2_eq}\\
{dv_2\over dr}&=&-2 {d\nu\over dr}\,h_2-\frac{2}{3}\,b^2\,e^{2 \nu} r^3\left(1+r {d\nu\over dr}\right)\left( ^{(0)} \rho ^{(0)} h\right)^2 \,, 
\nonumber \\ 
\label{dv2_eq} \\
\frac{\delta P_2 (r) }{ ^{(0)}\rho { } ^{(0)}h}&=&-h_2+\frac{1}{6 \pi }\,b^2\,e^{2 \nu}  r^2\left( ^{(0)} \rho ^{(0)} h\right)\,,\\
m_2&=&-e^{-2 \lambda}\,r\,h_2-\frac{2}{3}\,b^2\,e^{2 \left(\nu-\lambda\right)} r^5 \left( ^{(0)} \rho ^{(0)} h\right)^2\,.
\end{eqnarray}
The $\varepsilon_t \varepsilon_p$ order equations:
\begin{eqnarray}
d_2=-\frac{2}{3 }\,a\,b\,e^{\lambda+3 \nu} r^4 \left( ^{(0)} \rho ^{(0)} h\right)^2\,. \label{m2_eq}
\end{eqnarray}

Regular solutions of the master equations~(\ref{dm0_eq}), 
(\ref{dp0_eq}), (\ref{dh2_eq}), and (\ref{dv2_eq}) near the 
center of the star may be written as 
\beqn
m_0&=& r^3\left(m_{00}+m_{02} r^2\cdots\right) \,, \label{exp1} \\
 \frac{ \delta P_0(r)}{ ^{(0)}\rho { } ^{(0)}h}&=&h_{00}+h_{02} r^2+\cdots \,, \label{exp2} \\
h_2&=& r^2\left(h_{20}+h_{22} r^2+\cdots\right) \,, \\
v_2&=& r^4\left(v_{20}+v_{22} r^2+\cdots\right)\,, \label{exp4} \
\eeqn
where $m_{00}$, $m_{02}$, $h_{00}$, $h_{02}$, 
$h_{20}$, $h_{22}$, $v_{20}$, and $v_{22}$ are 
expansion coefficients. In this expansion solution, we may 
obtain a unique regular solution if values of $h_{00}$ and 
$h_{20}$ are given. In the present situation, a value of 
$h_{20}$ is determined by the boundary condition at infinity, 
which will be argued in the next paragraph. In order to 
determine a value of $h_{00}$, on the other hand, an extra 
condition is required. To determine the extra condition, in 
this study, we consider the following two distinct situations: 
(1) the baryon rest-mass density at the center of the star 
keeps constant when magnetic fields are imposed. (2) the 
total baryon rest-mass of the star keeps constant when magnetic 
fields are imposed. The situation (1) is realized by the 
condition of $h_{00}=0$ (cf., equation~(\ref{def_Delta_rho_c})). 
As for the situation (2), we will argue in the next subsection. 

The solutions of the $\varepsilon_t^2$-order vacuum 
Einstein equations suitable for the exterior spacetime 
of the isolated star are analytically given by  
\beqn
m_0=m_0(R)={\rm const.} \,, \quad h_0=-{m_0(R)\over r-2M}\,, \label{out_sol0}
\eeqn
\beqn
h_2=D Q_2^2(y)\,, \quad 
v_2&=&-{2 D\over \sqrt{y^2-1}}Q_2^1(y) \,, \label{out_sol2}
\eeqn
where $\displaystyle y= {r\over M}-1$ and, $Q_l^m$ and $D$ are the 
associated Legendre function of the second kind and 
a constant, respectively (cf., e.g., 
Refs.~\cite{Ioka2004,yoshidaa,thorne}). 
At the surface of the star, the external solutions, given by 
equations (\ref{out_sol0}) and (\ref{out_sol2}), are matched 
to the internal  solutions obtained by integrating equations 
(\ref{dm0_eq}), (\ref{dp0_eq}), (\ref{dv2_eq}), and 
(\ref{dh2_eq}) from the center of the star outwards with 
the boundary conditions given in 
equations~(\ref{exp1})--(\ref{exp4}). The physically 
acceptable solutions for the whole spacetime may 
then be obtained. 

\subsection{Global quantities characterizing the magnetized star}

In order to investigate properties of equilibrium solutions of the 
magnetized star, global quantities are used in the following 
discussion. For equilibrium states of the magnetized star, the 
total baryon rest-mass $\widetilde{M}^*$, the internal thermal 
energy $\widetilde{E}_{\rm int}$, and the electromagnetic 
energy $\widetilde{E}_{\rm EM}$ may be defined as
\beqn
&&\widetilde{M}^*=\int\rho\gamma\sqrt{-g}\,d^3x\,, \label{total_baryon}\\
&&\widetilde{E}_{\rm int}=\int\rho\varepsilon\gamma\sqrt{-g}\,d^3x\,,\\
&&\widetilde{E}_{\rm EM}={1\over 8\pi}\int B^\mu B_\mu \gamma \sqrt{-g}\,d^3x\ \label{magnetic}\,, 
\eeqn
(cf., e.g., Ref.~\cite{kiuchi}).

For the unperturbed spherical star, the gravitational mass 
$M$, the total baryon rest-mass $M^*$, and the internal 
thermal energy $E_{\rm int}$ may be given by
\beqn
&&M=M_r(R)\,, \\
&&M^*=4\pi \int_0^R {\,}^{(0)}\rho e^{\lambda} r^2 dr \,, \\ 
&&E_{\rm int}=4\pi \int_0^R {\,}^{(0)}\rho{\,}^{(0)}\varepsilon e^{\lambda} r^2 dr\,, 
\eeqn
where $R$ denotes the circumferential radius of the star 
determined by the condition ${\,}^{(0)}P(R)=0$.  The 
gravitational potential energy $W$ for the unperturbed 
star may be defined by
\beqn
|W|=M^*+E_{\rm int}-M\,. 
\eeqn

The $O(\epsilon_t^2)$ magnetic effects on the gravitational mass 
$\widetilde{M}$, the total baryon rest-mass 
$\widetilde{M}^*$, and the internal thermal energy 
$\widetilde{E}_{\rm int}$ may, respectively, be given by
\beqn
&&\Delta M = \epsilon_t^2\,m_0(R) \,, \\ 
&&\Delta M^*=4\pi \epsilon_t^2 \int_0^R  {\,}^{(0)}\rho e^{\lambda} r^2 
\nonumber \\ && \quad\quad \times 
\left({d\ln  {\,}^{(0)}\rho\over d  {\,}^{(0)}P}\delta P_0+{e^{2\lambda}m_0\over r}\right) dr
\,, \\ 
&&\Delta E_{\rm int}=4\pi \epsilon_t^2 \int_0^R  {\,}^{(0)}\rho  {\,}^{(0)}\varepsilon e^{\lambda} r^2 
\nonumber \\ &&  \quad\quad \times
\left({d\ln\left( {\,}^{(0)}\rho  {\,}^{(0)}\varepsilon\right)\over d {\,}^{(0)}P}\delta P_0+{e^{2\lambda}m_0\over
r} \right)dr\,.
\eeqn
As mentioned in the previous subsection, we study the 
sequences of equilibrium states of the magnetized star 
characterized by the fixed total baryon rest-mass. Thus, 
the condition of $\Delta M^*=0$ is used to determine 
values of $h_{00}$ in equation~(\ref{exp2}), which are 
related to $O(\epsilon_t^2)$ changes in the central 
density of the star $\Delta\rho_c$, given by
\beqn
\Delta\rho_c=\epsilon_t^2\, \left.{d{}^{(0)}\rho\over d{}^{(0)}P}\right|_{r=0} {}^{(0)}\rho(0) {}^{(0)}h(0)\, h_{00} \,. 
\label{def_Delta_rho_c}
\eeqn

The electromagnetic energy $E_{\rm EM}$ is decomposed as 
\beqn
E_{\rm EM}=E_{\rm EM}^{(p)}+E_{\rm EM}^{(t)} \,, 
\eeqn
where $E^{(p)}_{\rm EM}$ and $E_{\rm EM}^{(t)}$ are the poloidal 
and toroidal magnetic-field energies, respectively, given by
\beqn
&&E_{\rm EM}^{(p)}=\epsilon_p^2\, {1\over 3} \, a^2  
\int_0^R r^2 e^{4 \nu-\lambda} 
\nonumber \\ && \times 
\left[ \left\{ r {d\over dr}\left({\,}^{(0)}\rho {\,}^{(0)}h\right)+2 \left({\,}^{(0)}\rho {\,}^{(0)}h\right)
   \left(r {d\nu\over dr}+1\right)\right\}^2\right.  \nonumber \\
&& \quad\quad\quad\quad
 \left. +2  e^{2 \lambda} \left({\,}^{(0)}\rho {\,}^{(0)}h\right)^2 \right] \,dr + O\left(\varepsilon_p^4\right)\,, \\
&&E_{\rm EM}^{(t)}=\epsilon_t^2\,{1\over 3} \, b^2 \int_0^R  e^{\lambda+2\nu}
r^4\left({\,}^{(0)}\rho {\,}^{(0)}h\right)^2 \,dr \nonumber \\
&&\quad\quad\quad\quad + O\left(\varepsilon_t^4\right) \,.
\eeqn

Multipole moments of the star may characterize the equilibrium 
star globally. The constant of integration $D$ appearing in the 
exterior solution given in equation~(\ref{out_sol2}) 
is related to the mass quadrupole moment $\Delta Q$, given by
\begin{equation}
\Delta Q = \epsilon_t^2\, {8\over 5} \, M^3 D \,, 
\end{equation}
(cf., e.g., Refs.~\cite{ioka,thorne}).

Deformation of the surface of the star due to the magnetic
stress also characterizes equilibrium solutions of the 
magnetized star. The surface of the star is defined by the 
algebraic equation $P(r)={\,}^{(0)}P(r)+\epsilon_t^2 \delta P_0(r)
+\epsilon_t^2 \delta P_2(r) P_2(\theta)+O\left(\varepsilon_p^2\right)=0$. 
Thus, the $O(\epsilon_t^2)$ radial displacement of the fluid 
elements on the stellar surface, $\Delta r$, is given by
\beqn
\Delta r&=&\left(\Delta r\right)_0+\left(\Delta r\right)_2 P_2  \nonumber \\
&=&-\epsilon_t^2\, \left(\delta P_0(R)+\delta P_2(R) P_2\right){dr \over d{\,}^{(0)}P}(R) \,. 
\label{def_Dr}
\eeqn
$(\Delta r)_0$ may be interpreted as an average change 
in the radius of the star induced by the magnetic effects. 
The degree of the quadrupole surface deformation due 
to the magnetic stress is well described by the ellipticity 
$e^*$, given by
\beqn
e^*=-\epsilon_t^2\,{3\over 2} \left[ {\left(\Delta r\right)_2 \over R}+v_2(R)-h_2(R) \right] \,, 
\eeqn
where $e^*$ is defined as a relative difference between the 
equatorial and polar circumference radii of the star (cf., e.g., 
Refs.~\cite{Ioka2004,thorne}). Thus, $e^* <0$ ($e^* >0$) 
means that the star is prolate (oblate).

An important quantity of magnetized objects is the total 
magnetic helicity ${\cal H}$, which is a conserved quantity 
in ideal magnetohydrodynamics defined by
\beqn
{\cal H} = \int H^0 \sqrt{-g}\,d^3x\,, 
\label{Def_H}
\eeqn
where $H^0$ is the time component of the magnetic helicity 
four-current $H^\mu$, defined by 
\beqn
H^\mu = -{1\over 2} \epsilon^{\mu\nu\alpha\beta} A_\nu F_{\alpha\beta}\,. 
\label{Def_Hvector}
\eeqn
We may confirm that the magnetic helicity ${\cal H}$ is a conserved 
quantity in ideal magnetohydrodynamics as follows. Taking the the 
divergence of equation~(\ref{Def_Hvector}) yields 
\beqn
\nabla_\mu H^\mu=-{1\over 2}*F^{\mu\nu}F_{\mu\nu} \,,
\eeqn
where $*F_{\mu\nu}$ is the Hodge dual of the Faraday tensor 
$F_{\mu\nu}$. We then have $\nabla_\mu H^\mu=0$ if the 
perfect conductivity condition $F_{\mu\nu}u^\nu=0$ is satisfied. 
For the present model, the total magnetic helicity is explicitly 
written as
\beqn
{\cal H}&=&{16\pi\over 3} \, \epsilon_t \epsilon_p ab \int_0^R e^{\lambda+3 \nu} r^4 \left({\,}^{(0)}\rho {\,}^{(0)}h\right)^2 \,dr \,, 
\eeqn
where we use the vector potential $A_\mu$, given by 
\beqn
A_\mu =\left(0,\epsilon_t b \, e^{\lambda+\nu}  r^2 {\,}^{(0)}\rho {\,}^{(0)}h \cos\theta ,0,\epsilon_p \psi \right) \,. 
\eeqn
The dimensionless magnetic helicity, defined by ${\cal H}_M
 = {\cal H}/M^2$, is used when its numerical value is shown. The
magnetic helicity is a measure of the net twist of a magnetic-field
configuration. Thus, the magnetic helicity vanishes for purely
poloidal fields and for purely toroidal fields. Some experiments and
numerical computations show an interesting fact that the total
magnetic helicity is likely to be conserved even when the resistivity
cannot be ignored~\cite{ Braithwaite2004,Hsu2002}. If this fact is 
retained for the neutron star formation process, the total magnetic
helicity has to be approximately conserved during the formation
process.

\section{Numerical Results}\label{sec:result}

In this section, we give some numerical examples of the star 
including mixed poloidal-toroidal magnetic fields to examine the 
magnetic effects on the stellar structure. For the one-parameter 
equations of state (\ref{Def_one_p_EOS}), we use the polytrope 
and the gamma-law equation of state, respectively, given by 
\begin{eqnarray}
P&=&\kappa \, \rho^{1+{1\over n}} \,, \label{EOS:polytrope} \\
\varepsilon&=&{1\over \Gamma -1} {P\over\rho} \,, 
\label{EOS:gamma-law}
\end{eqnarray} 
where $\kappa$, $n$, $\Gamma$ are constants. The constant 
$n$ is called the polytrope index. The constant $\Gamma$ 
stands for the adiabatic index, defined by
\begin{eqnarray}
\Gamma = \left({\partial \ln P\over\partial \ln\rho} \right)_{\rm ad} \,, 
\end{eqnarray}
where ``ad'' means that the derivative is evaluated along an 
adiabatic process curve. A general relativistic version of the 
Schwarzschild discriminant $A$ for the background star may 
be defined by
\begin{eqnarray}
A = {1\over {}^{(0)} \rho  {}^{(0)}h}{d\left({}^{(0)}\rho+{}^{(0)}\rho {}^{(0)}\varepsilon\right)\over dr}
-{1\over\Gamma\,  {}^{(0)}P}{d {}^{(0)} P\over dr} \,. 
\end{eqnarray}
Following Ipser and Lindblom~\cite{ipser}, we employ a definition 
of the general relativistic Brunt-V\"ais\"al\"a frequency $N$, given by
\begin{eqnarray}
N^2=-g A \,, 
\end{eqnarray}
where $g$ is an effective gravitational acceleration, defined by
\begin{eqnarray}
g=-e^{2\nu-2\lambda}{1\over  {}^{(0)} \rho  {}^{(0)}h}{d{}^{(0)}P\over dr}\,. 
\end{eqnarray}
The sign of the discriminant $A$ therefore determines whether or 
not a stellar medium is stable against convection. A stellar medium, 
where $d{}^{(0)}\rho/dr\le 0$ and $d{}^{(0)} P/dr\le 0$ are fulfilled, is 
convectively unstable if $A>0$. If we assume that $\Gamma$ is 
given by 
\begin{eqnarray}
{1\over\Gamma} = {n\over n+1} +\tilde{\delta} \,, 
\end{eqnarray}
where $\tilde{\delta}$ is a constant, then the Schwarzschild 
discriminant can be written by 
\begin{eqnarray}
A = - {\tilde{\delta}\over {}^{(0)} h}\,{d\ln {}^{(0)} P\over dr} \,. 
\label{A_eq}
\end{eqnarray}
For a star characterized by the equations of state given in 
equations~(\ref{EOS:polytrope}) and (\ref{EOS:gamma-law}), thus, 
a stable stratification of the fluid density is realized if the following 
condition is fulfilled,  
\begin{eqnarray}
\tilde{\delta} < 0 \,, \quad {\rm or} \quad 
\Gamma > {n+1 \over n} \,. \label{Condition:stable_stratification}
\end{eqnarray}
For the isentropic case, we have $\Gamma=\left(n+1\right)/n$ 
(or $\tilde{\delta}=0$) and the stellar medium is marginally stable 
against convection. 

As argued in the last section, the conditions~(\ref{Condition:fluid}) 
have to be satisfied at the surface of star in order for the magnetic 
field to vanish there. These conditions are automatically fulfilled if 
$n>1$. In this study, we consider the case of $n=1.05$ only as an 
example of equations of state for ``neutron star-like'' models. 
Note that the $n=1$ polytrope is frequently used to study neutron 
stars. For the models with $n=1$, however,  the 
conditions~(\ref{Condition:fluid}) cannot be satisfied at the surface 
of the star as long as equation~(\ref{def:F_Psi}) is assumed. 

As for the adiabatic index, we choose two cases, 
$\Gamma=(n+1)/n \approx 1.95238$ and $\Gamma=2.05$. 
The former is a non-stratified case and the latter a stably stratified 
case (cf. equation~(\ref{Condition:stable_stratification})). As 
discussed in, e.g., Refs.~\cite{yoshidaa,reisenegger2008}, 
magnetic buoyancy instability can be weakened in a stably stratified 
stellar medium. As shown numerically by Mitchell et al.~\cite{mitchell}, 
the stable stratification will play a crucial role in order for large-scale 
magnetic fields inside the star to survive for a sufficiently long time 
(cf., also, Ref.~\cite{reisenegger2008}). In fact, the neutron star 
core is expected to be strongly stably stratified due to a smooth 
composition gradient~\cite{reisenegger}. Note that equations of state 
similar to those of the present study are employed in 
Ref.~\cite{yoshidaa}. 

\begin{figure*}[b]
\epsfxsize=3.5in
\hspace{-0.2cm}\vspace{0.8cm}\epsffile{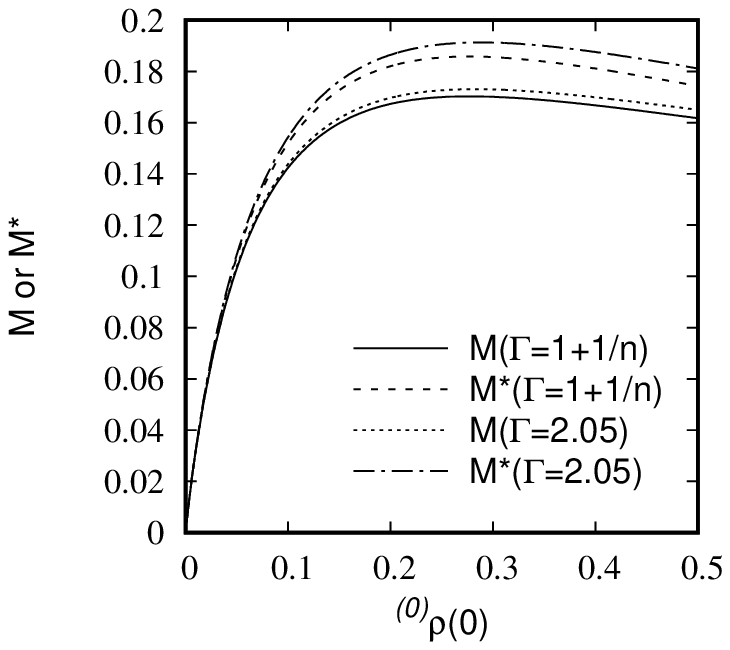}
\caption{Gravitational mass $M$ and baryon mass $M^*$, 
given as functions of the central density ${}^{(0)}\rho(0)$. 
All the quantities are given in units of $\kappa=1$.}
\label{f1}
\end{figure*}
For the background stars considered in this study, we 
plot in Figure \ref{f1} the gravitational mass~$M$ and 
the baryon rest-mass~$M^*$ as functions of the central 
density~${}^{(0)}\rho(0)$. Throughout this paper, units 
of $\kappa=1$ are used when numerical results are 
shown. The maximum gravitational mass is 
$M \approx 0.17028$ and $0.17307$ for the stars with 
$\Gamma=(n+1)/n$ and $2.05$, respectively. When the 
magnetized stars associated with the condition of 
$\Delta\rho_c=0$ are considered, the effects of the 
magnetic field on the stellar structure are examined 
for the background stars given in Figure \ref{f1}. 
When the the magnetized stars associated with the 
condition of $\Delta M^*=0$ are considered, we focus 
on the particular background stars with $M/R = 0.1$ 
and $0.2$. The compactness of $M/R = 0.2$ is typical 
for neutron stars. In Table \ref{tab1}, some global and 
physical quantities for these background stars are 
tabulated. Note that all the background stars given 
in Table \ref{tab1} are dynamically stable against radial 
collapse because values of their gravitational mass are 
less than those of the maximum one. 
\begin{table*}
\centering
\begin{minipage}{140mm}
\caption{\label{tab1}
Global and physical quantities for the background stars 
in units of $\kappa=1$. }
\begin{tabular}{cccccc}
\hline\hline
$\Gamma$ & $M/R$ & $^{(0)}\rho(0)$ & $M$ & $M^*$ & $E_{\rm{int}}/\left| W \right|$   \\
\hline
1.95238 & 0.100000&0.0625033&0.116229&0.122150&0.429344\\
              & 0.200000&0.252634 &0.170052&0.185529&0.601263\\
2.05000 & 0.100000&0.0620501&0.116627&0.122976&0.389322\\
              & 0.200000&0.244169 &0.172482&0.190336&0.543018\\
\hline
\end{tabular}
\end{minipage}
\end{table*}
\begin{figure*}[b]
\epsfxsize=1.8in
\leavevmode
\epsffile{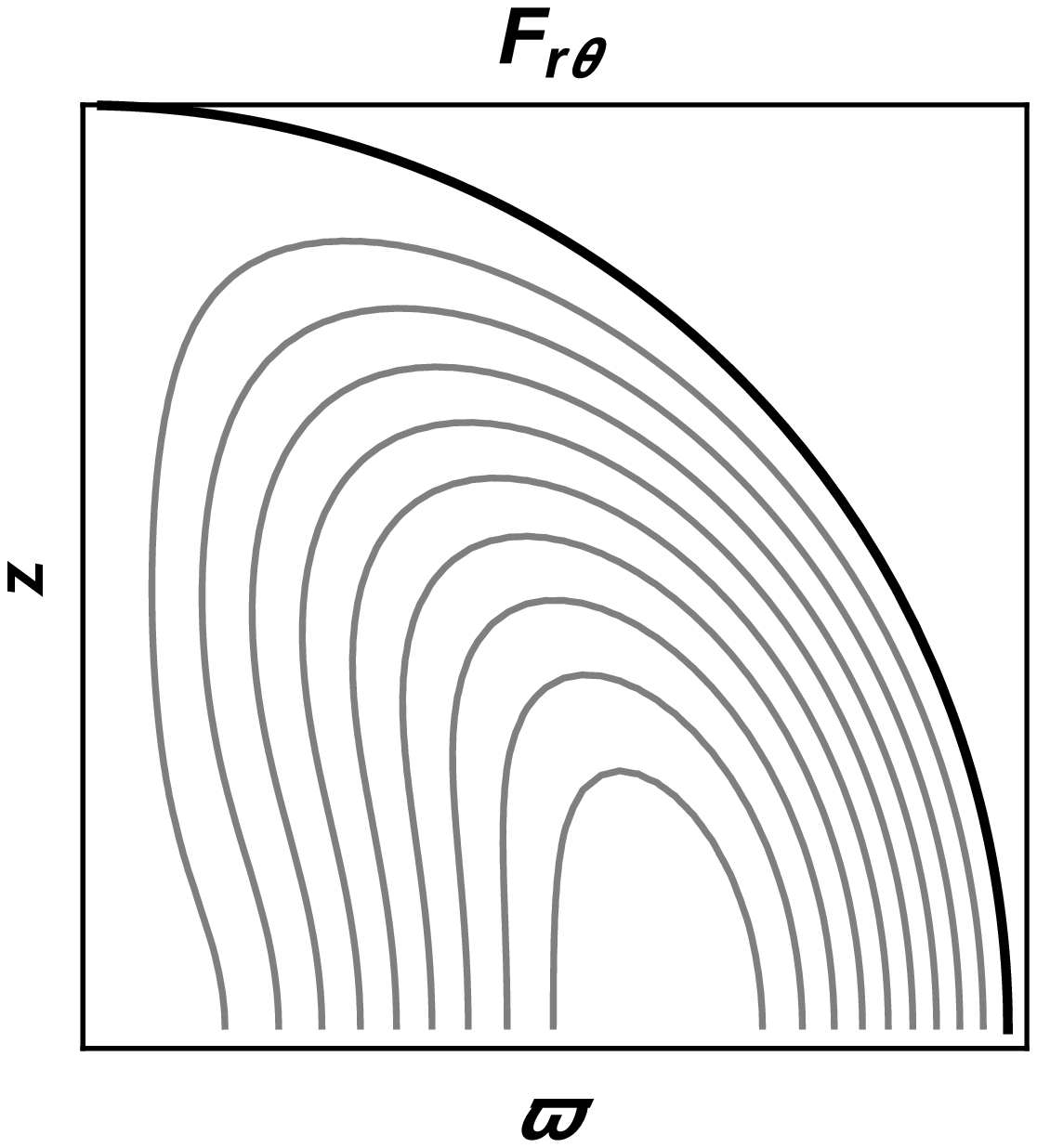}
\epsfxsize=1.8in
\leavevmode
\hspace{-0.2cm}\epsffile{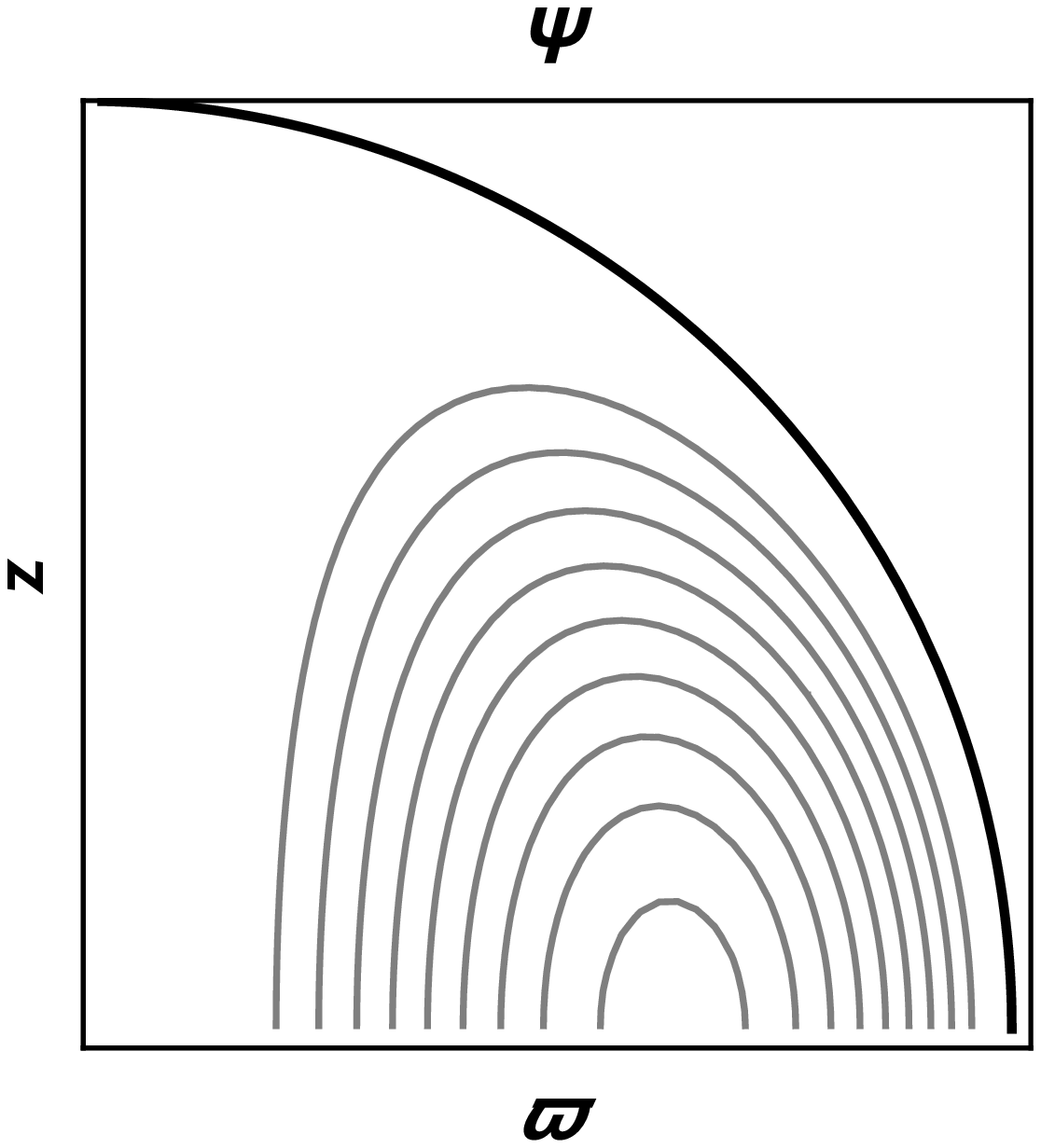}
\caption{Equi--$F_{r\theta}$ contours (Left) and 
equi--$\Psi$ contours (Right) on the meridional cross 
section for the model with $\Gamma=2.05$ and 
$M/R=0.2$. Here, $z$ and $\varpi$ are defined by 
$z=r\cos\theta$ and $\varpi=r\sin\theta$, respectively. 
The thick quarter circle shows the surface of the star, 
on which $F_{r\theta}=0$ and $\Psi=0$ are required 
by the boundary condition.}
\label{f3}
\end{figure*}

The distribution of magnetic fields is completely determined 
by the two arbitrary functions, $\Psi(w)$ and $K(w)$, with $w$ 
being the function of the background quantities. For the two  
arbitrary functions, we have assumed equations (\ref{def:F_K}) 
and (\ref{def:F_Psi}) in this paper.  In Figure \ref{f3}, we 
give the profiles of magnetic fields: the toroidal magnetic 
field $F_{r\theta}$ and the poloidal flux function $\Psi$ for 
the background star with $\Gamma=2.05$ and $M/R=0.2$. 
The right of Figure~\ref{f3} shows how lines of the 
magnetic force on the meridional cross section behave 
because an equi--$\Psi$ line corresponds to a line of the 
magnetic force. Note that there is no magnetic field 
outside the star in the model constructed in this study as 
mentioned before. 

For investigating the effects of magnetic fields on the 
structure of the star, it is helpful to introduce the 
quantities that represent typical strength 
of the magnetic field of the star. The norms of the 
toroidal and poloidal magnetic fields are, respectively, 
given by 
\begin{eqnarray}
\left| ^{(t)}B\right|^2 &=& \varepsilon_t^2  b^2 r^2\sin^2\theta \, e^{2\nu} \left({}^{(0)}\rho ^{(0)}h \right) ^2  \,, \\
\left| ^{(p)}B\right|^2 &=&\varepsilon_p^2 a^2\, 4 e^{4\nu}  \left({}^{(0)}\rho  ^{(0)} h \cos\theta \right)^2 \nonumber \\
&&+e^{4\nu-2\lambda}\sin^2\theta \nonumber \\
&&\times \left\{ r {d\over dr}\left( ^{(0)}\rho  ^{(0)}h \right)
\right. \nonumber \\ && \left. 
\quad +2 \left( ^{(0)}\rho  ^{(0)}h\right) \left(r {d\nu\over dr}+1\right)\right\}^2 \,. 
\end{eqnarray}
The maximum value of the norm of the toroidal 
magnetic field is then given by
\begin{eqnarray}
\left| ^{(t)}B_{\rm max}\right|^2 = \varepsilon_t^2  b^2 \, {\rm max}\left[ r^2 \, e^{2\nu} \left({}^{(0)}\rho ^{(0)}h \right) ^2 \right] \,, 
\end{eqnarray}
where ${\rm max}\left[f(r)\right]$ means the maximum 
value of the function $f(r)$. The maximum value of 
the norm of the poloidal magnetic field is mostly 
attained at the center of the star. Thus, the norm of 
the poloidal magnetic field at the center of the star, 
given by 
\begin{eqnarray}
\left| ^{(p)}B_c\right|^2 =\varepsilon_p^2 a^2\, 4 e^{4\nu(0)}  \left({}^{(0)}\rho(0)  ^{(0)} h(0)\right)^2  \,, 
\end{eqnarray}
may be used as a representative value of the 
strength of the poloidal magnetic field. By using 
these values of the norms, we may obtain the two 
dimensionless quantities representing 
magnetic-field strength, given by 
\begin{eqnarray}
 ^{(t)}{\cal R}_M = {\left| ^{(t)}B_{\rm max}\right|^2 R^4\over 4 M^2} \,, 
 ^{(p)}{\cal R}_M = {\left| ^{(p)}B_c\right|^2 R^4\over 4 M^2}  \,, 
\end{eqnarray}
which are as large as the ratios of the toroidal and 
the poloidal magnetic energies to the gravitational 
energy, respectively. The perturbations due to the 
magnetic field given in equation~(\ref{comp:faraday}) 
basically depend on the dimensionless smallness 
parameters $\varepsilon_t$ and $\varepsilon_p$, 
which can be arbitrarily set depending on the desired 
magnetic field-strength as long as 
$1 \gg \varepsilon_t \gg \varepsilon_p$. This 
arbitrariness in perturbation quantities is then 
removed by using the two dimensionless quantities 
$^{(t)}{\cal R}_M$ and $^{(p)}{\cal R}_M$ when 
numerical results are shown in this paper. 

\begin{figure*}[b]
\epsfxsize=3.5in
\hspace{-0.2cm}\vspace{0.8cm}\epsffile{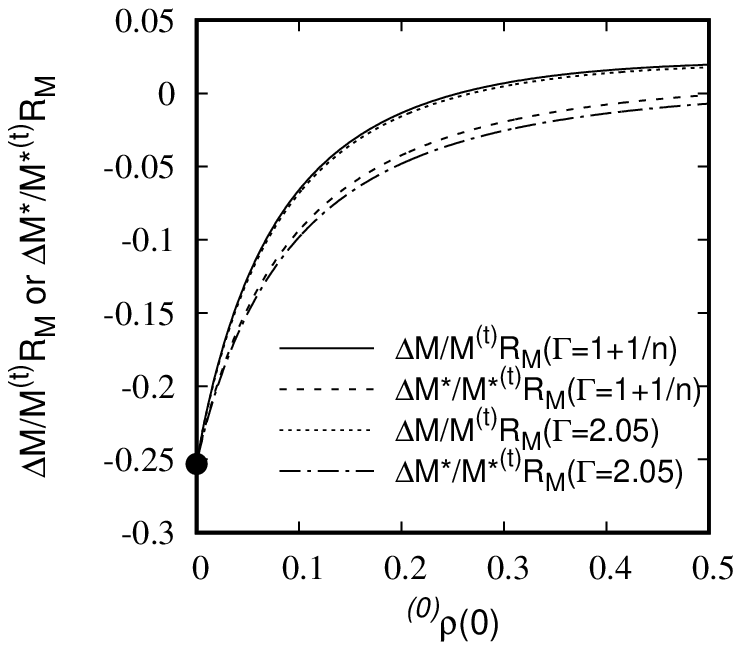}
\caption{Normalized nondimensional changes in the 
gravitational mass $\Delta M/M{}^{(t)}{\cal R}_M$ and 
the baryon rest-mass $\Delta M^*/M^*{}^{(t)}{\cal R}_M$, 
given as functions of the central density of the 
background star ${}^{(0)}\rho(0)$. The solid circle 
indicates the result obtained within the framework of 
Newton's dynamics and theory of gravity (cf., Appendix).}
\label{f4}
\end{figure*}

\begin{figure*}[b]
\epsfxsize=3.5in
\hspace{-0.2cm}\vspace{0.8cm}\epsffile{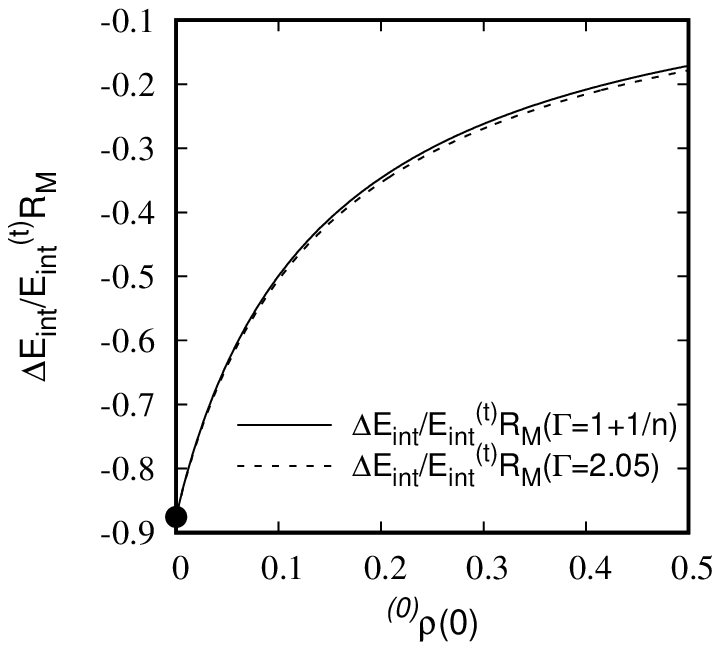}
\caption{Normalized nondimensional changes in the internal 
thermal energy $\Delta E_{\rm int}/E_{\rm int}{}^{(t)}{\cal R}_M$, 
given as functions of the central density of the background star 
${}^{(0)}\rho(0)$. The solid circle indicates the result obtained 
within the framework of Newton's dynamics and theory of 
gravity (cf., Appendix).}
\label{f5}
\end{figure*}

\begin{figure*}[b]
\epsfxsize=3.5in
\hspace{-0.2cm}\vspace{0.8cm}\epsffile{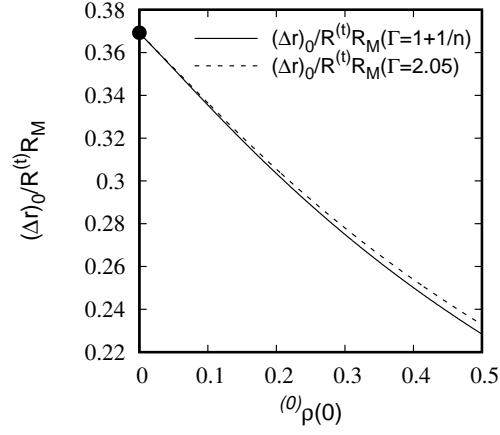}
\caption{Normalized nondimensional changes in the mean 
radius $(\Delta r)_0/R{}^{(t)}{\cal R}_M$, given as functions of 
the central density of the background star ${}^{(0)}\rho(0)$. The 
solid circle indicates the result obtained within the framework 
of Newton's dynamics and theory of gravity (cf., Appendix).}
\label{f6}
\end{figure*}

\begin{figure*}[b]
\epsfxsize=3.5in
\hspace{-0.2cm}\vspace{0.8cm}\epsffile{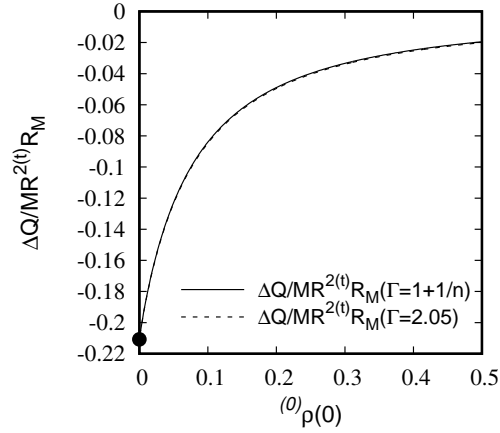}
\caption{Normalized nondimensional changes in the mass 
quadrupole moment $\Delta Q/MR^2{}^{(t)}{\cal R}_M$, given as 
functions of the central density of the background star 
${}^{(0)}\rho(0)$. The solid circle indicates the result obtained 
within the framework of Newton's dynamics and theory of 
gravity (cf., Appendix).}
\label{f7}
\end{figure*}

\begin{figure*}[b]
\epsfxsize=3.5in
\hspace{-0.2cm}\vspace{0.8cm}\epsffile{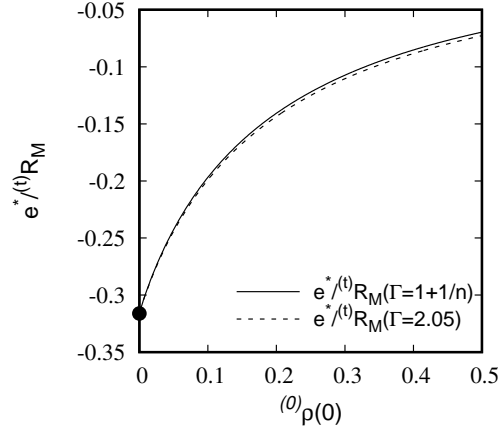}
\caption{Normalized changes in the ellipticity $e^*/{}^{(t)}{\cal R}_M$, 
given as functions of the central density of the background star 
${}^{(0)}\rho(0)$. The solid circle indicates the result obtained 
within the framework of Newton's dynamics and theory of 
gravity (cf., Appendix).}
\label{f8}
\end{figure*}

\begin{figure*}[b]
\epsfxsize=3.5in
\hspace{-0.2cm}\vspace{0.8cm}\epsffile{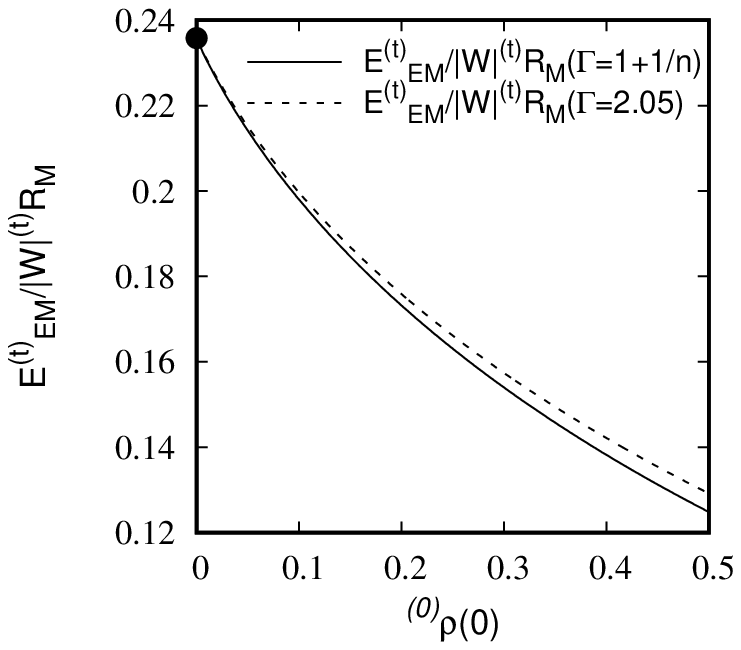}
\caption{Normalized nondimensional changes in the toroidal 
magnetic energy $E_{\rm EM}^{(t)}/\left| W \right| {}^{(t)}{\cal R}_M$, 
given as functions of the central density of the background star 
${}^{(0)}\rho(0)$. The solid circle indicates the result obtained 
within the framework of Newton's dynamics and theory of 
gravity (cf., Appendix).}
\label{f9}
\end{figure*}

\begin{figure*}[b]
\epsfxsize=3.5in
\hspace{-0.2cm}\vspace{0.8cm}\epsffile{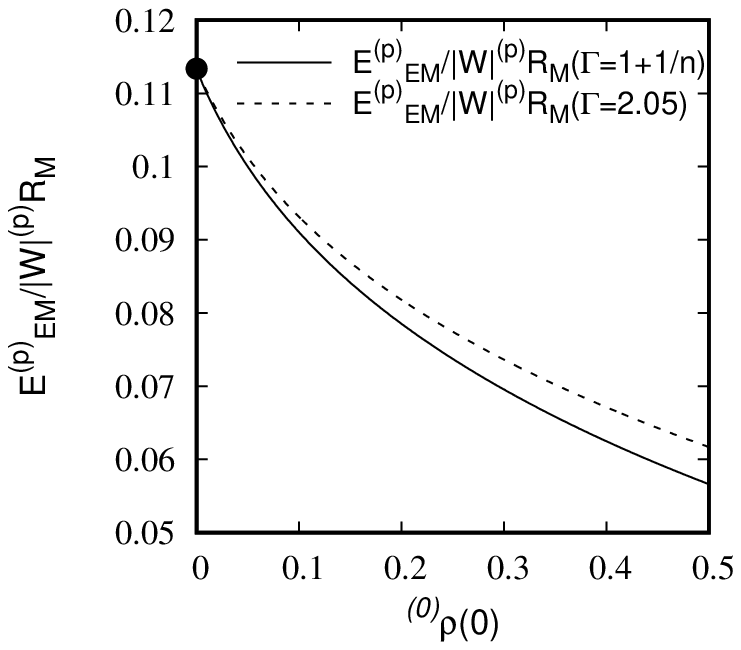}
\caption{Normalized nondimensional changes in the poloidal 
magnetic energy $E_{\rm EM}^{(p)}/\left| W \right| {}^{(p)}{\cal R}_M$, 
given as functions of the central density of the background star 
${}^{(0)}\rho(0)$. The solid circle indicates the result obtained 
within the framework of Newton's dynamics and theory of 
gravity (cf., Appendix).}
\label{f10}
\end{figure*}

\begin{figure*}[b]
\epsfxsize=3.5in
\hspace{-0.2cm}\vspace{0.8cm}\epsffile{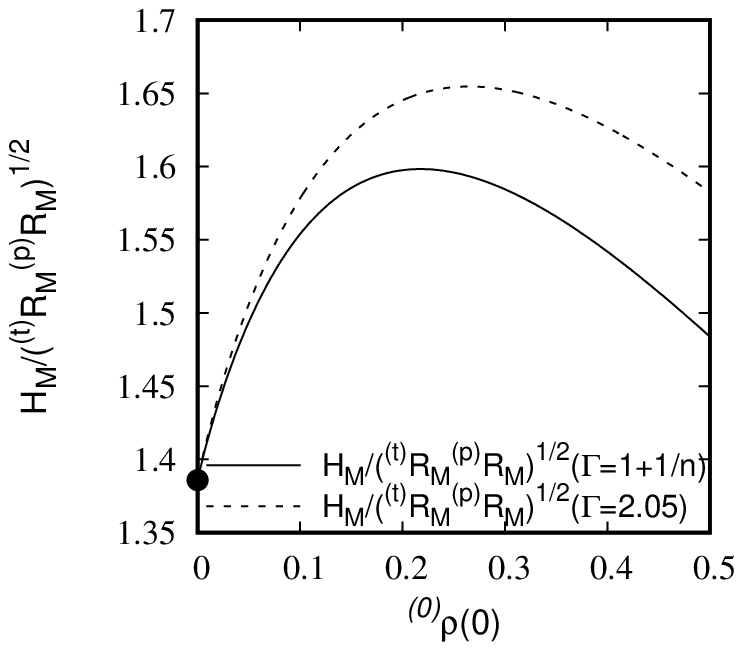}
\caption{Normalized changes in the nondimensional magnetic 
helicity ${\cal H}_M/\sqrt{{}^{(t)}{\cal R}_M {}^{(p)}{\cal R}_M}$, 
given as functions of the central density of the background star 
${}^{(0)}\rho(0)$. The solid circle indicates the result obtained 
within the framework of Newton's dynamics and theory of 
gravity (cf., Appendix).}
\label{f11}
\end{figure*}

First, we examine properties of the magnetized stars 
obtained under the condition of $\Delta\rho_c=0$, i.e., 
their central densities are kept constant when the 
magnetic fields are imposed. In Figure~\ref{f4}, 
normalized nondimensional changes in the 
gravitational mass $\Delta M \left/M{}^{(t)}{\cal R}_M \right.$ 
and the baryon rest-mass 
$\Delta M^* \left/ M^*{}^{(t)}{\cal R}_M \right.$ are plotted 
as functions of the central density of the background star 
${}^{(0)}\rho(0)$. In Figures~\ref{f5} through~\ref{f11}, we 
plot, as functions of the central density of the background 
star ${}^{(0)}\rho(0)$, normalized nondimensional changes 
in the internal thermal energy 
$\Delta E_{\rm int} \left/ E_{\rm int}{}^{(t)}{\cal R}_M \right.$, 
the mean radius 
$\left(\Delta r\right)_0 \left/ R{}^{(t)}{\cal R}_M \right.$, 
the mass quadrupole moment 
$\Delta Q \left/ MR^2{}^{(t)}{\cal R}_M \right.$, the ellipticity 
$e^* \left/ {}^{(t)}{\cal R}_M \right.$, the toroidal magnetic 
energy 
$E_{\rm EM}^{(t)} \left/ |W|{}^{(t)}{\cal R}_M \right.$, and 
the poloidal magnetic energy 
$E_{\rm EM}^{(p)} \left/ |W|{}^{(p)}{\cal R}_M \right.$, the 
magnetic helicity 
${\cal H}_M \left/ \sqrt{{}^{(t)}{\cal R}_M{}^{(p)}{\cal R}_M}\right.$, 
respectively. In Figures~\ref{f4} through \ref{f11}, the solid 
circles on the vertical axis indicate the results obtained by 
the calculation based on Newtonian magnetohydrodynamics 
and Newton's theory of gravity (cf., Appendix). In these figures, 
we see that the present general relativistic results in the 
Newtonian limit (the limit of ${}^{(0)}\rho(0)\to 0$) are nicely 
agreement with those obtained by the Newtonian calculations. 
This fact serves as a useful consistency check of our 
numerical code. 

Properties of the magnetized stars with $\Delta\rho_c=0$ 
observed in Figures~\ref{f4} through~\ref{f11} are summarized 
as follows: The results for the models with $\Gamma= 2.05$ are 
little different from those for the models with 
$\Gamma= (n+1)/n\approx 1.95238$ (cf. Figures~\ref{f4} through~\ref{f11}). 
This implies that small stratification has little effect on the equilibrium 
structure of the magnetized stars. The imposition of the 
toroidal magnetic fields results in a decrease in the total 
baryon rest-mass, i.e., $\Delta M^*<0$. Due to the imposition 
of the toroidal magnetic fields, values of the gravitational mass 
decrease, i.e., $\Delta M<0$, for the background stars with 
${}^{(0)}\rho(0) \lessapprox 0.255$ while they increase, i.e., 
$\Delta M>0$, for the background stars with 
${}^{(0)}\rho(0) \gtrapprox 0.255$ (cf. Figure~\ref{f4}). The 
mean radius of the star $\left(\Delta r\right)_0$ increases 
when the toroidal magnetic field is imposed (cf. Figure~\ref{f6}).
The values of the mass quadrupole moment $\Delta Q$ and 
the ellipticity $e^*$ are negative, which reflects the fact that 
the star is prolate (cf. Figures~\ref{f7} and \ref{f8}). The prolate 
deformation is typical for stars containing dominant toroidal 
magnetic fields (cf., e.g., Refs~\cite{kiuchi,frieben}). 

\begin{table*}
\centering
\begin{minipage}{192mm}
\caption{\label{tab2}
Global and physical quantities.}
\begin{tabular}{cccccccccc}
\hline\hline
$(\Gamma,M/R)$ & $\displaystyle{\Delta\rho_c\over{}^{(0)}\rho(0){}^{(t)}{\cal R}_M}$ & 
$\displaystyle{\Delta M\over M{}^{(t)}{\cal R}_M}$ & $\displaystyle{\Delta E_{\rm int}\over E_{\rm int}{}^{(t)}{\cal R}_M}$ & 
$\displaystyle{(\Delta r)_0\over R{}^{(t)}{\cal R}_M}$ &$\displaystyle{\Delta Q\over MR^2{}^{(t)}{\cal R}_M}$ & $\displaystyle{e^*\over{}^{(t)}{\cal R}_M}$ &
$\displaystyle{E_{\rm EM}^{(t)}\over|W|{}^{(t)}{\cal R}_M}$ & $\displaystyle{E_{\rm EM}^{(p)}\over|W|{}^{(p)}{\cal R}_M}$& 
$\displaystyle{{\cal H}_M\over \sqrt{{}^{(t)}{\cal R}_M{}^{(p)}{\cal R}_M}}$\\
\hline
(1.95238,0.1)&0.2424&$1.565\times 10^{-2}$&$-0.2413$&                         0.3048&$-0.1105$                    &$-0.2294$&0.2097 &$9.747\times 10^{-2}$&1.512\\
(1.95238,0.2)&0.8077&$2.338\times 10^{-2}$&     0.4077&$-1.607\times 10^{-2}$&$-3.946\times 10^{-2}$&$-0.1213$&0.1626&$7.349\times 10^{-2}$&1.596\\
(2.05,0.1)      &0.2457&$1.646\times 10^{-2}$&$-0.2397$&                         0.3056&$-0.1116$                    &$-0.2316$&0.2110 &$9.904\times 10^{-2}$&1.527\\
(2.05,0.2)      &0.5677&$2.257\times 10^{-2}$&     0.2043& $8.193\times 10^{-2}$&$-4.134\times 10^{-2}$&$-0.1271$&0.1672&$7.792\times 10^{-2}$&1.654\\
\hline
\end{tabular}
\end{minipage}
\end{table*}
Next, we examine properties of the magnetized stars obtained 
under  the condition of $\Delta M^*=0$, i.e., their total baryon 
rest-masses are kept constant when the magnetic fields are imposed. 
Table \ref{tab2} lists global and physical quantities characterizing   
the magnetized stars with $\Delta M^*=0$; the changes in the 
central density $\Delta\rho_c$, the gravitational mass $\Delta M$, 
the internal thermal energy $\Delta E_{\rm int}$, the mean radius 
$\left(\Delta r\right)_0$, the mass quadrupole moment $\Delta Q$, 
the ellipticity $e^*$, the toroidal magnetic energy $E_{\rm EM}^{(t)}$, 
the poloidal magnetic energy $E_{\rm EM}^{(p)}$, and the magnetic 
helicity ${\cal H}$. In this table, all the quantities are normalized 
to be nondimensional, as given in the first row. 

Properties of the magnetized star with $\Delta M^*=0$ 
observed in Table~\ref{tab2} are summarized as follows: The 
imposition of the toroidal magnetic fields results in an increase 
in the central density, i.e.,  $\Delta\rho_c>0$. The values of the mass 
quadrupole moment $\Delta Q$ and the ellipticity $e^*$ are 
negative, which reflects the fact that the star is prolate. These 
properties concerning $\Delta\rho_c$ and $\Delta Q$ are 
attributed to the magnetic hoop stress around the symmetry 
axis due to the toroidal magnetic field, which tends to make the 
star prolate like a rubber belt fastening around the waist of a star. 
The gravitational mass increase due to the imposition of the 
toroidal magnetic fields, i.e., $\Delta M>0$. 

Since the deformation of the star considered in this study is 
caused by toroidal magnetic fields only, even though poloidal 
magnetic fields make the deformation of the spacetime, the 
results obtained in this study can be compared with those 
obtained by Kiuchi and Yoshida~\cite{kiuchi}, who derived 
general relativistic stars having  purely toroidal magnetic 
fields with a non-perturbative  approach. Although weakly 
magnetized stars cannot be calculated with Kiuchi and 
Yoshida's method because of their non-perturbative 
approach, it is found that the present results are consistent 
with those obtained by Kiuchi and Yoshida~\cite{kiuchi} 
(Compare, e.g., Table~\ref{tab2} with Figure~6 of 
Ref.~\cite{kiuchi}). 

\section{Discussion}\label{sec:discussion}

In this study, as mentioned before, the general relativistic 
magnetized stars are constructed under the 
condition of $\varepsilon_p \ll \varepsilon_t \ll 1$, and 
the effects of magnetic fields are investigated within 
accuracy $O\left(\varepsilon_t \varepsilon_p\right)$. The terms 
higher than $O\left(\varepsilon_p^2\right)$ in the equations 
are then discarded. The deformation of the star and spacetime 
occurs in the $\varepsilon_t^2$-order, which is attributed to 
the magnetic effects due to the toroidal field. This deformation 
due to the toroidal magnetic field is the same as that of the 
weakly magnetized star with purely toroidal fields within 
accuracy $O\left(\varepsilon_t^2\right)$. Thus, the 
present results include those for the weakly magnetized 
star with purely toroidal fields. To our knowledge, such 
general relativistic magnetized stars having purely toroidal 
fields have been constructed with a perturbative approach 
for the first time. The $\varepsilon_t \varepsilon_p$-order 
effects appear in the deformation of the spacetime only. 
Therefore, the $\varepsilon_t \varepsilon_p$-order 
effects are general relativistic ones and disappear in 
Newton's dynamics and theory of gravity. Within the 
framework of Newtonian magnetohydrodynamics, in 
other words, the poloidal magnetic field does not affect 
the deformation of the star within accuracy 
$O\left(\varepsilon_t \varepsilon_p\right)$ (cf. Appendix). From 
a general relativistic point of view, an interesting fact is 
that  the $\varepsilon_t \varepsilon_p$-order effects violate 
the circularity conditions (cf. equation~(\ref{circularity_C})).  
As a result, the $r\varphi$-component of the metric 
$g_{r\varphi}$ appears inside the star (cf. 
equation~(\ref{m2_eq})). 

In this paper, we have shown that stationary and 
axisymmetric solutions of the magnetized star with 
mixed poloidal-toroidal fields may indeed be 
constructed within accuracy 
$O\left(\varepsilon_t\varepsilon_p\right)$. However, 
such stationary and axisymmetric solutions cannot 
be constructed if the $\varepsilon_p^2$-order effects 
on the structure of the star are included. The reason 
for this is the following: The $\varepsilon_p^2$-order 
equations are the same as those for the weakly 
magnetized star with purely poloidal fields within 
accuracy $O\left(\varepsilon_p^2\right)$. For the 
weakly magnetized star with purely poloidal fields, 
the poloidal flux function $\Psi$ has to satisfy the 
general relativistic version of the so-called 
Grad-Shafranov equation (cf., e.g., 
Ref.~\cite{konno}). In the present approximation, 
however, the flux function $\Psi$ has to be given by 
the arbitrary function of the background quantity  
$w=^{(0)}\rho ^{(0)}h e^{2\nu}r^2\sin^2\theta$ 
(cf. equations  (\ref{eq:Euler-phi}) and 
(\ref{eq:Euler-poloidal})). The flux function $\Psi$ 
given by the arbitrary function of $w$ does not, in 
general, fulfill the Grad-Shafranov equation. Therefore, 
the $\varepsilon_p^2$-order equations cannot be 
solved consistently with the lower-order equations. 
This implies that the weakly magnetized stars 
constructed in this study cannot be stationary and 
axisymmetric when the condition of 
$\varepsilon_p \ll \varepsilon_t$ is violated. 

After obtaining equilibrium models of stars, check of their 
stability is an  important issue because unstable solutions 
lose their physical meaning in the sense that they are not 
realized in nature. Since magnetized stars with purely 
toroidal fields are unstable, the present magnetized star 
models are indeed unstable when we set 
$\varepsilon_p=0$, which corresponds to the case of 
purely toroidal fields. As mentioned in Introduction, both 
a stable stratification of the fluid and poloidal magnetic 
fields act as stabilizing agents of the toroidal magnetic 
fields inside the star. The stably stratified stars with 
$1\gg\varepsilon_t \gg \varepsilon_p \ne 0$ constructed 
in this study are therefore possibly stable. As mentioned 
before, unfortunately, reliable and useful procedures for 
the diagnosis of the stability for the magnetized star have 
not yet been established. (For the moment, numerical 
simulations will be the most reliable way to check the 
stability, but they are tough work.) Although we are not 
sure that it is adaptive for the present magnetized star 
models, Braithwaite's stability condition, given in 
equation~(\ref{Braithwaite_SC}), is available to assess 
their stability. If a magnetized star characterized  
by $\Gamma=2.05$, $R\approx10$~km, 
$M\approx1.4 \,  M_\odot$, and 
$^{(t)}B_{\rm max} \approx 10^{15}$~G is considered, 
we have $^{(t)}{\cal R}_M\approx 5 \times 10^{-7}$ 
and $E^{(t)}_{\rm EM}/|W|\approx 8\times 10^{-8}$. 
We then obtain Braithwaite's stability condition for the 
model, given by 
\beq
 8\times 10^{-5}  \lesssim  {E_{\rm EM}^{(p)} \over E_{\rm EM}^{(t)}}
\lesssim 0.8 \,, 
\label{Braithwaite_SC_ex1}
\eeq
where 
$E_{\rm EM}^{(p)}/E_{\rm EM} \approx E_{\rm EM}^{(p)}/E_{\rm EM}^{(t)}$ 
is used, and $\tilde{a}\approx 10^3$ is assumed because 
the star is a stably stratified neutron star model. For the 
model considered, we have 
$E_{\rm EM}^{(p)}/E_{\rm EM}^{(t)}\approx 0.5 \, {\cal R}_M^{(p)}/{\cal R}_M^{(t)}$. 
Braithwaite's stability condition for the model then 
becomes
\beq
 1.6\times 10^{-4}  \lesssim  {{\cal R}_M^{(p)} \over {\cal R}_M^{(t)}}
\lesssim 1.6 \,. 
\label{Braithwaite_SC_ex2}
\eeq
Under the condition of $\varepsilon_p \ll \varepsilon_t \ll 1$, 
which is the basic assumption in this study, we can 
appropriately choose vales of 
${{\cal R}_M^{(p)} / {\cal R}_M^{(t)}}$ so as to satisfy 
the inequality given in 
equation~(\ref{Braithwaite_SC_ex2}), Braithwaite's stability 
condition for the model. Therefore, the present magnetized 
star models satisfying the inequality~(\ref{Braithwaite_SC_ex2}) 
are stable if Braithwaite's stability condition is properly adaptive 
for them. In order to examine stability properly, however, we 
have to  make stability analyses by using dynamical 
simulations or solving linear eigenvalue problems, which 
exceed the scope of this work and remain as future challenges. 

\section{Summary}\label{sec:summary}

We have constructed the stably stratified magnetized 
stars within the framework of general relativity. The 
effects of magnetic fields on the structure of the star and 
spacetime are treated as perturbations of  
non-magnetized stars. By assuming ideal 
magnetohydrodynamics and employing 
one-parameter equations of state, we derive 
basic equations for describing stationary and 
axisymmetric stably stratified stars containing  
magnetic fields whose toroidal components are much 
larger than the poloidal ones. A number of the polytropic 
models are numerically calculated to investigate 
basic properties of the effects of magnetic fields on 
the stellar structure. According to the stability result 
obtained by Braithwaite, which remains a matter of 
conjecture for general magnetized stars, certain of 
the magnetized stars constructed in this study 
are possibly stable.  

\section*{Acknowledgments}

This work was supported by the Grant-in-Aid for 
Scientific Research (C) No.~18K03606.

\appendix

\section{Newtonian analysis}\label{sec:apdixA}

In this appendix, we present the Newtonian version of the 
magnetized star considered in this study. The results of 
the Newtonian analysis can be used to compare to those 
of the general relativistic analysis in the Newtonian limit. 
We may then check consistency between them. Similar 
analyses to those given in this appendix are found in, e.g.,  
Refs.~\cite{sinha,asai}. 

Within the framework of Newtonian magnetohydrodynamics, 
the dynamics of perfectly conductive fluids may be 
described by the following equations:
\begin{equation}
\partial_t \,\rho+\nabla_a\left(\rho\, v^a\right)=0 \,,
\label{a1}
\end{equation}
\begin{equation}
\nabla_aB^a=0 \,,
\label{a2}
\end{equation}
\begin{equation}
\partial_t \,B^a=\nabla_b\left(v^a B^b-v^b B^a\right) \,,
\label{a3}
\end{equation}
\begin{eqnarray}
\left(\partial_t +v^b\nabla_b\right)v_a
&=&-{1\over \rho}\nabla_a p-\nabla_a\Phi
\nonumber \\ 
&+&{1\over 4\pi\rho} \left( B^b\nabla_b B_a-B^b\nabla_a B_b \right)
 \,, 
\label{a4}
\end{eqnarray}
\begin{equation}
\nabla^b\nabla_b \Phi=4\pi G \rho \,,
\label{a5}
\end{equation}
where $\rho$, $v^a$, $B^a$, $p$, and $\Phi$ are the 
mass density, the fluid velocity, the magnetic field, 
the pressure, and the gravitational potential, 
respectively. Here, $\nabla_a$ denotes the covariant 
derivative associated with the metric $g_{ab}$, and 
spatial indices are denoted by lower case Roman 
letters ($a, b, c, \cdots$). 

Following the assumptions given in 
Sec.~\ref{sec:formulation}, we assume that there is 
no fluid flow, i.e., 
\begin{equation}
v^a=0 \,, 
\label{def_v}
\end{equation}
and that the magnetized stars are stationary and 
axisymmetric. Therefore, physical quantities 
associated with the magnetized star are 
independent of the time coordinate $t$ and the 
azimuthal angle about the symmetry axis 
$\varphi$. Under the assumption of stationarity 
and axisymmetry, the magnetic fields $B^a$ 
may, in terms of two functions $B$ and 
$A_\varphi$ independent of $t$ and $\varphi$, 
be written by 
\begin{equation}
\quad B^a=B \varphi^a+\epsilon^{ab\varphi}\partial_b A_\varphi \,,
\label{def_B}
\end{equation}
where $\varphi^a$ denotes the rotational Killing 
vector,  $\epsilon^{abc}$ is the contravariant 
spatial Levi-Civita tensor, and $A_\varphi$ is 
the $\varphi$-component of the vector potential 
$A_a$ or the poloidal flux function. Due to the 
assumptions given in 
equations~(\ref{def_v}) and (\ref{def_B}), 
equations~(\ref{a1})--(\ref{a3}) are satisfied 
automatically, and the $\varphi$-component 
of equation~(\ref{a4}) becomes 
\begin{eqnarray}
{1\over 4\pi\rho} \left( B^b\nabla_b B_\varphi-B^b\nabla_\varphi B_b \right)
&=&{1\over 4\pi\rho} \epsilon^{bc\varphi}\partial_c A_\varphi\partial_b
 \left( B \, \varphi _\varphi  \right) \nonumber \\
 &=&0 \,. 
\label{Euler_phi_comp}
\end{eqnarray}
Therefore, the function $B$ has to be given 
in terms of an arbitrary function 
$K(A_\varphi)$ by 
\begin{equation}
B={K(A_\varphi)\over\varphi_\varphi} \,. 
\label{Def_B_fun}
\end{equation}
By using equation~(\ref{Def_B_fun}), we may 
rewrite the poloidal components of the Lorentz 
force term in equation~(\ref{a4}) as follows: 
\begin{eqnarray}
&&{1\over 4\pi\rho} \left( B^b\nabla_b B_C-B^b\nabla_C B_b \right)
=-{K\over4\pi\rho\,\varphi_\varphi}\partial_C K
\nonumber \\
&&\quad\quad\quad \quad
-{1\over4\pi\rho\sqrt{g}}\left( \partial_C A_\varphi\right)\left( \partial_2 B_1-\partial_1 B_2 \right) \,,  
\label{Euler_poloidal_comp}
\end{eqnarray}
where the index $C$ denote the poloidal 
indices, i.e., $C=1,2$, and $g$ means the 
determinant of the metric $g_{ab}$. If we 
make the same approximation as that used in 
Sec.~\ref{sec:formulation}, the first and the 
second terms in the right-hand side of 
equation~(\ref{Euler_poloidal_comp}) are 
$O\left(\varepsilon_t^2\right)$ and 
$O\left(\varepsilon_p^2\right)$, respectively. Under 
the assumption of 
$\varepsilon_p \ll \varepsilon_t \ll 1$, 
equation~(\ref{Euler_poloidal_comp}) 
becomes 
\begin{equation}
{1\over 4\pi\rho} \left( B^b\nabla_b B_C-B^b\nabla_C B_b \right)
=-{K\over4\pi\rho\,\varphi_\varphi}\partial_C K+O\left(\varepsilon_t\varepsilon_p\right) \,,  
\end{equation}
within accuracy $O\left(\varepsilon_t\varepsilon_p\right)$. 
In other words, similarly to the general 
relativistic case, poloidal magnetic fields 
do not affect the deformation of the star 
within accuracy $O\left(\varepsilon_t\varepsilon_p\right)$. 
The Euler equation then becomes 
\begin{equation}
{1\over \rho}\nabla_C p+\nabla_C\Phi
+{1\over8\pi\rho\,\varphi_\varphi}\partial_C K^2
+O\left(\varepsilon_t\varepsilon_p\right)=0 \,. 
\label{Euler2}
\end{equation}
This equation may be integrable if the following 
conditions for the three functions $p$, $K$, and 
$A_\varphi$, are assumed: 
\begin{equation}
p=p(\rho)\,, \quad K=K(\rho\varphi_\varphi)\,, 
\quad A_\varphi =A_\varphi(\rho\varphi_\varphi) \,.
\end{equation}
After giving the actual forms of the three functions 
$p$, $K$, and $A_\varphi$, we may then obtain 
the weakly magnetized star models with mixed 
poloidal and toroidal fields. 

In what follows, the spherical polar coordinates 
$(r,\theta,\varphi)$ are used in order to derive 
the master equations for the weakly 
magnetized stars with mixed poloidal-toroidal 
fields. The metric is then given by 
\begin{equation}
ds^2=dr^2+r^2d\theta^2+r^2\sin^2\theta d\varphi^2 \,.
\end{equation}
The rotational Killing vector $\varphi_a$ is given by
\begin{equation}
\varphi_a=(0,0,r^2\sin^2\theta) \,.
\end{equation}
Following the assumptions given in 
Sec.~\ref{sec:formulation}, we set the two arbitrary 
functions $K$ and $A_\varphi$ as follows: 
\begin{eqnarray}
K&=&{B_c \over \sqrt{2}\,\rho_c \alpha}\, \rho \, r^2 \sin^2\theta \,, 
\label{def_K_L} \\
A_\varphi&=&b \rho \, r^2\sin^2 \theta \,, 
\end{eqnarray}
where $B_c$,  $\rho_c $, and $\alpha$ are constants 
that are related to the magnetic field strength, density, 
and radius of the star, respectively, and $b$ is a constant. 
The absolute value of the magnetic field $B^a$ is then 
given by 
\begin{equation}
\sqrt{B^a B_a} = {B_c \over \sqrt{2}}\, \hat{\rho} \, \xi \sin\theta +O\left(\varepsilon_p^2\right) \,, 
\end{equation}
where the two dimension-less quantities 
$\hat{\rho}=\rho/\rho_c$ and $\xi=r/\alpha$ are 
introduced. By using equation~(\ref{def_K_L}), 
we may rewrite equation~(\ref{Euler2}) as 
\begin{eqnarray}
&&\nabla_a p=-\rho \nabla_a \Psi \,, 
\label{Euler3} \\
&&\Psi 
=\Phi +{1 \over 3}\,\Omega_A^2 r^2 \, \hat{\rho} \left\{1-P_2\left(\cos\theta\right)\right\} -\Phi_0 \,, 
\end{eqnarray}
where $\Omega_A$ is the Alfv\`en frequency, defined by 
$\displaystyle \Omega_A=\sqrt{ {B_c^2 \over 4\pi \rho_c \alpha^2}}$, 
and $\Phi_0$ is a constant. Since $\Omega_A = O\left(\varepsilon_t\right)$, 
within accuracy $O\left(\varepsilon_t\varepsilon_p\right)$, 
the function $\Psi$ fulfills the equation, given by 
\begin{eqnarray}
\nabla^a \nabla_a\Psi &=&4 \pi G \rho \nonumber \\
&+&{1 \over 3}\,\Omega_A^2 
\left\{ r^2{d^2\hat{\rho}_0\over dr^2}+6 r {d\hat{\rho}_0\over dr}+6 \hat{\rho}_0
\nonumber \right. \\
&& \left. -\left(r^2 {d^2\hat{\rho}_0\over dr^2}+6 r{d\hat{\rho}_0\over dr}\right) P_2\left(\cos\theta\right) \right\}
\nonumber  \\
&& +O\left(\varepsilon_p^2\right) \,, \label{nabla_Psi}
\end{eqnarray}
where $\hat{\rho}_0$ is the dimensionless density 
of the non-magnetized star normalized by its central 
value.  The function $\Psi$ may be expanded in 
terms of the parameter $\Omega_A$, and then written by 
\begin{eqnarray}
\Psi\left(r,\theta\right)&=&\Psi_0\left(r\right)
\nonumber \\ 
&-&2\alpha^2\Omega_A^2\left[\psi_0\left(r\right)+\psi_2\left(r\right)P_2\left(\cos\theta\right) \right]
\nonumber \\
&& +O\left(\varepsilon_p^2\right) \,, \label{exp_Psi}
\end{eqnarray}
where $\Psi_0$ means the function $\Psi$ for the 
non-magnetized star. Since we have $p=p(\rho)$ 
and equation~(\ref{Euler3}), the density $\rho$ is a 
function of $\Psi$. Therefore, the density $\rho$ 
may be also expanded in terms of the parameter 
$\Omega_A$, and then written by 
\begin{eqnarray}
\rho\left(r,\theta\right)&=&\rho_0\left(r\right)
\nonumber \\ 
&-& 2\alpha^2\Omega_A^2{d\rho_0 \over d\Psi_0}
\left[\psi_0\left(r\right)+\psi_2\left(r\right)P_2\left(\cos\theta\right) \right] 
\nonumber \\
&&+O\left(\varepsilon_p^2\right)
\,, \label{exp_rho}
\end{eqnarray}
where $\rho_0$ means the density $\rho$ for the 
non-magnetized star. Instituting 
equations~(\ref{exp_Psi}) and (\ref{exp_rho}) into 
equation~(\ref{nabla_Psi}), we obtain 
\begin{eqnarray}
\alpha^2 \nabla^a \nabla_a \Psi_0\left(r\right) = 4 \pi G \alpha^2 \rho_0\left(r\right) \,, 
\label{beq0}
\end{eqnarray}
\begin{eqnarray}
{1\over \xi^2}{d \over d\xi}\left(\xi^2\, {d \psi_0\over d\xi}\right)
&=&k(\xi) \psi_0
\nonumber \\
&-&{1 \over 6r^2}{d\over dr}\left(r^2{d\over dr}\left(r^2 \hat{\rho}_0\right)\right) \,, 
\label{beq1}
\end{eqnarray}
\begin{eqnarray}
{1\over \xi^2}{d \over d\xi}\left(\xi^2\, {d \psi_2\over d\xi}\right)
&=& \left(k\left(\xi\right)+{6\over \xi^2} \right)\psi_2
\nonumber \\
&+&{1 \over 6}\left( r^2 {d^2\hat{\rho}_0\over dr^2}+6 r {d\hat{\rho}_0\over dr} \right) \,, 
\label{beq2}
\end{eqnarray}
where 
\begin{equation}
k\left(\xi\right)=4 \pi G \alpha^2{d\rho_0 \over d\Psi_0} \,. 
\label{def_k}
\end{equation}
In order to solve the three ordinary differential 
equations~(\ref{beq0})--(\ref{beq2}), boundary 
conditions at the center and surface of the star 
are necessary. We require that physical quantities 
are regular near the center of the star and that 
values of the central density are independent of 
the magnetic field-strength. At the center of the 
star, therefore, we have 
\begin{eqnarray}
&&{d\Psi_0\over d\xi}\left(0\right)=0  \,, \quad 
\psi_0\left(0\right)=0  \,, \quad  {d\psi_0\over d\xi}\left(0\right)=0  \,,  \nonumber \\ 
&&\psi_2\left(0\right)=0  \,, \quad  {d\psi_2\over d\xi}\left(0\right)=0  \,. 
\label{BC_cent} 
\end{eqnarray}
To determine the boundary condition at the surface 
of the star, we need the equation of the surface 
of the star, given by
\begin{equation}
r=R\left(1+\delta\zeta\right)  +O\left(\varepsilon_p^2\right) \,, 
\end{equation}
where $R$ is the radius for the 
non-magnetized star. The dimensionless displacement 
$\delta\zeta$ is given by 
\begin{eqnarray}
\delta\zeta={2\alpha^2\Omega_A^2\over \displaystyle R{d\Psi_0\over dr}(R)}
\left[\psi_0\left(\xi_1\right)+\psi_2\left(\xi_1\right)P_2\left(\cos\theta\right) \right] \,,
\end{eqnarray}
where $\xi_1=R/\alpha$. The displacement $\delta\zeta$ 
is used to evaluate the ellipticity of the 
surface of the star $e^*$, defined by 
\begin{eqnarray}
e^*&=&{R\left(1+\delta\zeta\left(\pi/2\right)\right)-R\left(1+\delta\zeta\left(0\right)\right)\over R} \nonumber \\
&=&-{3\over 2}2\alpha^2\Omega_A^2{1\over \displaystyle R{d\Psi_0\over dr}(R)}\psi_2\left(\xi_1\right) \,.
\end{eqnarray}
The gravitational potentials inside and outside the star within 
accuracy $O\left(\varepsilon_t^2\right)$ are, respectively, given by 
\begin{eqnarray}
\Phi &=&\Psi_0\left(r\right)+c_0 \nonumber \\
&-&2\alpha^2\Omega_A^2\left[c_{1,0}+\psi_0\left(\xi\right)+\psi_2\left(\xi\right)P_2\left(\cos\theta\right) \right]
\nonumber \\
&-&{1 \over 3}\,\Omega_A^2 r^2 \, \hat{\rho} \left\{1-P_2\left(\cos\theta\right)\right\} \,, 
\end{eqnarray}
and 
\begin{equation}
\Phi =-{\kappa_0\over \xi}
-2\alpha^2\Omega_A^2\left[{\kappa_{1,0}\over \xi}+{\kappa_{1,2}\over \xi^3}P_2\left(\cos\theta\right) \right] \,, 
\end{equation}
where $c_0$, $c_{1,0}$, $\kappa_0$, $\kappa_{1,0}$ and 
$\kappa_{1,2}$ are constants. Since the gravitational 
potential and its derivative have to be continuous at 
the surface of the star, we obtain the following relations: 
\begin{equation}
\kappa_0=\xi_1^2\partial_\xi\Psi_0\left(R\right) \,,  \ c_0=-\Psi_0\left(R\right)-\xi_1\partial_\xi\Psi_0\left(R\right)\,, 
\label{bc_potential0}
\end{equation}
\begin{equation}
-c_{1,0}=-{\kappa_{1,0}\over \xi_1}+\psi_0\left(\xi_1\right)\,, \quad {\kappa_{1,2}\over \xi_1^3}=\psi_2\left(\xi_1\right) \,, 
\label{bc_potential1}
\end{equation}
\begin{eqnarray}
&&{\kappa_{1,0}\over \xi_1^2}=-{d\psi_0\over d\xi}(\xi_1)-{1 \over 6}\, \xi_1^2 {d \hat{\rho}\over d\xi}\left(\xi_1\right)\,, 
\nonumber \\
&&3{\kappa_{1,2}\over \xi_1^4}=-{d\psi_2\over d\xi}(\xi_1)+{1 \over 6}\, \xi_1^2 {d \hat{\rho}\over d\xi}\left(\xi_1\right) \,. 
\label{bc_potential2}
\end{eqnarray}
For equations~(\ref{beq0}) and (\ref{beq1}), therefore, 
boundary conditions are not imposed at the surface of 
the star, and instead the constants characterizing the 
gravitational potential are determined through 
equation~(\ref{bc_potential0}) and 
\begin{eqnarray}
\kappa_{1,0}&=&-\xi_1^2{d\psi_0\over d\xi}(\xi_1)-{1 \over 6}\, \xi_1^4 {d \hat{\rho}\over d\xi}\left(\xi_1\right)\,, 
\nonumber \\ 
c_{1,0}&=&-\psi_0(\xi_1)-\xi_1{d\psi_0\over d\xi}(\xi_1)-{1 \over 6}\, \xi_1^3 {d \hat{\rho}\over d\xi}\left(\xi_1\right) \,,
\end{eqnarray}
As for equation~(\ref{beq2}), the boundary condition 
at the surface of the star is given by
\begin{equation}
3\psi_2\left(\xi_1\right)+\xi_1{d\psi_2\over d\xi}\left(\xi_1\right)={1 \over 6}\, \xi_1^3 {d \hat{\rho}\over d\xi}\left(\xi_1\right) \,. 
\end{equation}
The constant related to the mass quadrupole moment 
$\kappa_{1,2}$ is determined through 
\begin{equation}
\kappa_{1,2}=\xi_1^3 \psi_2\left(\xi_1\right) \,. 
\end{equation}

Following the assumptions given in 
Sec.~\ref{sec:formulation}, we assume the 
polytropic equation of state, given by
\begin{eqnarray}
p=\kappa\rho^{1+{1\over n}}  \,, 
\end{eqnarray}
where $\kappa$ and $n$ are constants. Introducing 
the Lane-Emden function $\Theta$, then, we 
may write $\rho$ and $p$ as 
\begin{eqnarray}
\rho=\rho_c \Theta^n \,, \quad  p=p_c\Theta^{n+1} \,, 
\label{rho_p_Theta}
\end{eqnarray}
where $\rho_c$ and $p_c$ are values of the 
density and pressure at the center of the star, 
respectively. The central pressure value $p_c$ 
is given in terms of $\rho_c$, $\kappa$, and $n$ 
by $p_c=\kappa\rho_c^{1+{1\over n}}$. 
Equation~(\ref{beq0}) is rewritten by 
\begin{eqnarray}
{1\over\xi^2}{\partial\over\partial\xi}\left(\xi^2{\partial\over\partial\xi} \widehat{\Psi} \right) =\Theta^n \,, 
\label{poisson}
\end{eqnarray}
where $\widehat{\Psi}$ is the dimensionless 
quantity, defined by
\begin{eqnarray}
\widehat{\Psi}={\Psi\over4\pi G\rho_c\alpha^2} \,. 
\end{eqnarray}
From equation~(\ref{Euler3}), inside the star, we obtain 
\begin{eqnarray}
\widehat{\Psi}=-\Theta+C \,, 
\label{rel_Psi_Theta}
\end{eqnarray} 
where $C$ is a constant, and we set 
\begin{eqnarray}
\alpha=\sqrt{{(n+1)K\rho_c^{{1\over n}}\over 4 \pi G\rho_c}}=\sqrt{{(n+1)p_c\over 4 \pi G\rho_c^2}} \,. 
\end{eqnarray} 
Substituting equation~(\ref{rel_Psi_Theta}) into 
equation~(\ref{poisson}) yields the Lane-Emden 
equation, given by 
\begin{eqnarray}
{d^2\Theta\over d\xi^2}+{2\over\xi}{d\Theta\over d\xi}=-\Theta^n \,. 
\label{LM_eq}
\end{eqnarray}
At the center of the star, the boundary conditions for 
equation~(\ref{LM_eq}) are, due to 
equations~(\ref{BC_cent}) and (\ref{rho_p_Theta}), 
given by
\begin{eqnarray}
\Theta=1\,, \quad {d\Theta\over d\xi}=0\,, \quad {\rm at}\ \xi=0 \,. 
\end{eqnarray}
The function $k(\xi)$, defined in equation~(\ref{def_k}), 
is rewritten by  
\begin{eqnarray}
k\left(\xi\right)=-n\, \Theta^{n-1} \,. 
\end{eqnarray}
Now that we obtain the complete set of equations 
for the weakly magnetized star with mixed 
poloidal-toroidal fields within the framework of 
Newtonian magnetohydrodynamics, which are 
composed of equations~ (\ref{LM_eq}), (\ref{beq1}), 
and (\ref{beq2}), we may construct the magnetized 
stars. 

In order to investigate properties of equilibrium 
solutions of the magnetized star, global quantities 
are frequently used. The mass of the star $M$ is 
given by
\begin{eqnarray}
M&=&2\pi \int \rho r^2\sin\theta dr d\theta \,, \nonumber \\
&=&M_0\left(1+{\Delta M\over M_0}\right)+O\left(\varepsilon_t^2\right) \,,
\end{eqnarray}
where $M_0$ is the mass of the non-magnetized 
star, given by
\begin{eqnarray}
M_0=-4\pi\alpha^3 \rho_c \left. \xi_1^2{d\Theta\over d\xi}\right|_{\xi=\xi_1} \,, 
\end{eqnarray}
and $\displaystyle{\Delta M\over M_0}$ is the normalized 
dimensionless change in the mass of the star, 
given by
\begin{eqnarray}
{\Delta M\over M_0}={\displaystyle 
{\Omega_A^2\over 2\pi G\rho_c}\left\{\left. {d \psi_0\over d\xi}\right|_{\xi=\xi_1}
+{1 \over 6}\left. \xi_1^2{d\hat{\rho}_0\over d\xi}\right|_{\xi=\xi_1} \right\} 
\over \displaystyle \left. {d\Theta\over d\xi}\right|_{\xi=\xi_1} }\,. 
\end{eqnarray}
The internal thermal energy of the star $E_{\rm int}$ is 
given by
\begin{eqnarray}
E_{\rm int}&=&2\pi \int \rho \varepsilon r^2\sin\theta dr d\theta \,, \nonumber \\
&=&\left(E_{\rm int}\right)_0\left(1+{\Delta E_{\rm int}\over 
\left(E_{\rm int}\right)_0}\right)+O\left(\varepsilon_t^2\right) \,,
\end{eqnarray}
where $\left(E_{\rm int}\right)_0$ is the internal thermal 
energy of the non-magnetized star, given by
\begin{eqnarray}
\left(E_{\rm int}\right)_0={4\pi\over \Gamma-1}{4\pi G\rho_c^2\alpha^5\over n+1} 
\int_0^{\xi_1} \Theta^{n+1} \xi^2 d\xi \,,
\end{eqnarray}
and $\displaystyle{\Delta E_{\rm int}\over \left(E_{\rm int}\right)_0}$ 
is the normalized dimensionless change in the internal 
thermal energy of the star, given by
\begin{eqnarray}
{\Delta E_{\rm int}\over \left(E_{\rm int}\right)_0}={\displaystyle 
{\Omega_A^2 \over 4\pi G\rho_c}2\left(n+1\right)\int_0^{\xi_1} \Theta^n \psi_0\left(r\right) \xi^2 d\xi
\over \displaystyle \int_0^{\xi_1} \Theta^{n+1} \xi^2 d\xi } \,. 
\end{eqnarray}
Here, the gamma-law equation of state, given in 
equation~(\ref{EOS:gamma-law}), is used. 
An average change in the radius of the star 
$\left( \Delta r\right)_0$, defined in 
equation~(\ref{def_Dr}), is given by 
\begin{eqnarray}
{\left( \Delta r\right)_0 \over R}=-2\Omega_A^2{\psi_0\left(\xi_1\right) \over
\displaystyle 4\pi G\rho_c  \xi_1\left.{d\Theta\over d\xi}\right|_{\xi=\xi_1}} \,. 
\end{eqnarray}
The ellipticity associated with the surface shape of 
the star $e^*$ is explicitly given by
\begin{eqnarray}
e^* = 3 \Omega_A^2{\psi_2\left(\xi_1\right) \over
\displaystyle 4\pi G\rho_c  \xi_1\left.{d\Theta\over d\xi}\right|_{\xi=\xi_1}} \,. 
\end{eqnarray}
The mass quadrupole moment of the star 
$\Delta Q$ is, in terms of $\kappa_{1,2}$, 
given by
\begin{eqnarray}
{\Delta Q\over M_0R^2}=-{2\alpha^2\Omega_A^2\alpha^3\kappa_{1,2}\over \alpha\kappa_0 \alpha^2 \xi_1^2} \,. 
\end{eqnarray}
The toroidal magnetic energy $^{(t)} E_{\rm EM}$ and 
the poloidal magnetic energy $^{(p)} E_{\rm EM}$ are, 
respectively, defined by 
\begin{eqnarray}
&&^{(t)} E_{\rm EM}={1\over 8\pi}\int B^\varphi B_\varphi dV \,,\nonumber \\
&&^{(p)} E_{\rm EM}={1\over 8\pi}\int B^CB_C dV \,. 
\end{eqnarray}
Then, the ratios of the toroidal and the poloidal magnetic 
energies to the unperturbed gravitational energy of the star 
$\displaystyle{{}^{(t)}E_{\rm EM}\over \left(\left| W \right|\right)_0}$ and 
$\displaystyle{{}^{(p)}E_{\rm EM}\over \left(\left| W \right|\right)_0}$ are, 
respectively, given by 
\begin{eqnarray}
{{}^{(t)}E_{\rm EM}\over \left(\left| W \right|\right)_0}
={\displaystyle
{ \Omega_A^2 \over 3\cdot 4\pi \rho_c G}\int_0^{\xi_1} \hat{\rho}^2 \xi^4 d\xi
 \over \displaystyle -\int_0^{\xi_1} \hat{\rho}\widehat{\Phi} \xi^2 d\xi  } \,,
\end{eqnarray}
and 
\begin{eqnarray}
 &&{{}^{(p)}E_{\rm EM}\over \left(\left| W \right|\right)_0}
={b^2\over 3\cdot 2\pi \alpha^2 4\pi G} \nonumber \\
&&\quad\quad \times {\displaystyle \int_0^{\xi_1}  \left\{   
 \left( \xi{d\hat{\rho}\over d\xi}+2\hat{\rho} \right)^2+2 \rho^2 \right\} \xi^2 d\xi \over 
 \displaystyle  
 -\int_0^{\xi_1} \hat{\rho}\widehat{\Phi} \xi^2 d\xi  } \,, 
\end{eqnarray}
where the unperturbed gravitational energy of the star $\left(\left| W \right|\right)_0$ is given by  
\begin{eqnarray}
\left(\left| W \right|\right)_0&=&-{1\over 2}\int \rho_0\Phi_0 dV \,,\nonumber \\
&=&-2\pi \alpha^5 \rho_c^2 4\pi G \int_0^{\xi_1} \hat{\rho}_0\widehat{\Phi}_0 \xi^2 d\xi \,. 
\end{eqnarray}
The dimensionless magnetic helicity of the star ${\cal H}_M$ is given by 
\begin{eqnarray}
{\cal H}_M&=&{{\cal H}\over G M_0^2} \,,\nonumber \\
&=& {\displaystyle 
{b B_c  \over \sqrt{2}\,3\pi G \alpha^2 \rho_c}  \int_0^{\xi_1} \hat{\rho}^2 \xi^4 d\xi \over 
\displaystyle  \left( \left. \xi_1^2{d\Theta\over d\xi}\right|_{\xi=\xi_1}\right)^2 } \,, 
\end{eqnarray}
where ${\cal H}$ is the magnetic helicity of the star, defined by 
${\cal H}=\int A_a B^a dV$. 

Since the effects of magnetic fields on the structure 
of the star are treated as perturbations of the non-magnetized 
star, the solutions constructed in this study are inherently 
independent of the magnetic-field strength. However, the 
representation for the global and physical quantities defined before 
are dependent on the magnetic-field strength. In order to remove 
their field-strength dependence, following the treatment used 
in Sec.~\ref{sec:result}, we introduce the two dimensionless 
quantities representing magnetic-field strength, given by 
\begin{eqnarray}
{}^{(t)}{\cal R}_M &=&{^{(t)}B_{\rm max}^2 R^4\over 4GM^2} \,, \nonumber \\
&=&{\left( {\rm max}\left[ \hat{\rho} \xi \right]\right)^2 \over \displaystyle
8 \left( \left. {d\Theta\over d\xi}\right|_{\xi=\xi_1} \right)^2}{\Omega_A^2 \over 4\pi G \rho_c} \,, \\
{}^{(p)}{\cal R}_M &=&{^{(p)}B_c^2 R^4\over 4GM^2} \,, \nonumber \\
&=&{b^2 \over \displaystyle G \left(4\pi\right)^2 \alpha^2 \left( \left. {d\Theta\over d\xi}\right|_{\xi=\xi_1} \right)^2} \,, 
\end{eqnarray}
where $^{(t)}B_{\rm max}$ is the maximum absolute value 
of the toroidal magnetic field inside the star, 
${\rm max}[f]$ means the maximum value of the function $f$, 
and $^{(p)}B_c$ is the absolute value of the poloidal 
magnetic field at the center of the star. ${}^{(t)}{\cal R}_M$ 
and ${}^{(p)}{\cal R}_M$ are as large as the ratios of the 
toroidal and the poloidal magnetic energies to the gravitational 
energy, respectively. 

\begin{table*}
\centering
\begin{minipage}{192mm}
\caption{\label{tabApendix}
Global and physical quantities. }
\begin{tabular}{ccccccccc}
\hline\hline
$(\Gamma,M/R)$ &  
$\displaystyle{\Delta M\over M{}^{(t)}{\cal R}_M}$ & $\displaystyle{\Delta E_{\rm int}\over E_{\rm int}{}^{(t)}{\cal R}_M}$ & 
$\displaystyle{(\Delta r)_0\over R{}^{(t)}{\cal R}_M}$ &$\displaystyle{\Delta Q\over MR^2{}^{(t)}{\cal R}_M}$ & $\displaystyle{e^*\over{}^{(t)}{\cal R}_M}$ &
$\displaystyle{E_{\rm EM}^{(t)}\over|W|{}^{(t)}{\cal R}_M}$ & $\displaystyle{E_{\rm EM}^{(p)}\over|W|{}^{(p)}{\cal R}_M}$& 
$\displaystyle{{\cal H}_M\over \sqrt{{}^{(t)}{\cal R}_M{}^{(p)}{\cal R}_M}}$\\
\hline
(2.05,0)      &$-0.2532$&$-0.8757$& $0.3692$&$-0.2107$&$-0.3161$&0.2358&0.1134&1.386\\
\hline
\end{tabular}
\end{minipage}
\end{table*}
In this appendix, we numerically obtain the magnetized star 
model assuming that $n=1.05$ and $\Gamma=2.05$, 
which is the Newtonian version of the weakly magnetized 
general relativistic star model calculated in this study. 
Table \ref{tabApendix} lists global and physical quantities characterizing 
the magnetized stars constructed within the framework of Newtonian 
magnetohydrodynamics; the changes in the mass $\Delta M$, 
the internal energy $\Delta E_{\rm int}$, the mean radius 
$(\Delta r)_0$, the mass quadrupole moment $\Delta Q$, the ellipticity 
$e^*$, the toroidal magnetic energy $E_{\rm EM}^{(t)}$, the 
poloidal magnetic energy $E_{\rm EM}^{(p)}$, and the dimensionless 
magnetic helicity ${\cal H}_M$. In this table, all the quantities are 
normalized to be nondimensional, as given in the first row.

\end{document}